\documentclass[aps,prl,twocolumn,superscriptaddress]{revtex4-2}
\usepackage{color}
\usepackage{bm,bbm}
\usepackage{amsmath}
\usepackage{amsfonts}
\usepackage{amssymb}
\usepackage{array}
\usepackage{diagbox}
\usepackage{tikz}
\usepackage{extarrows}
\usepackage{txfonts}
\usepackage{graphicx}
\usepackage{url}
\usepackage[colorlinks=true, urlcolor=blue, linkcolor=blue, citecolor=blue, pdftex]{hyperref}
\usepackage{braket}

\newcommand{\beqn}{\begin{eqnarray}}
\newcommand{\eeqn}{\end{eqnarray}}

\newcommand{\linkedsup}[2]{\hyperlink{#1}{\textsuperscript{#2}}}
\newcommand{\footanchor}[1]{\hypertarget{#1}{}}

\newenvironment{myfootnotes}{
  \footnotesize
  \begin{list}{}{\setlength{\labelwidth}{1.2em}%
                 \setlength{\leftmargin}{2.5em}%
                 \setlength{\itemindent}{0em}%
                 \setlength{\parsep}{0em}%
                 \setlength{\itemsep}{0em}%
                 \setlength{\topsep}{0.5em}}%
}{%
  \end{list}%
}
\newcommand{\myfootnote}[3]{\item[\hfill#1\hspace{0.3em}]\footanchor{#2}#3}

\begin{document}

\title{Gossamer Superconductivity in Moir\'e WSe$_2$ Bilayer}

\author{Hui-Ke Jin\linkedsup{fn:common}{*}$^{,}$\linkedsup{fn:jin}{$\dagger$}}
\affiliation{School of Physical Science and Technology, ShanghaiTech University, Shanghai 201210, China}

\author{Guangyue Ji\linkedsup{fn:common}{*}}
\affiliation{International Center for Quantum Materials, Peking University, Beijing 100871, China}
\affiliation{Beijing Key Laboratory of Quantum Devices, Peking University, Beijing, 100871, China}

\author{Zhan Wang}
\affiliation{Beijing National Laboratory for Condensed Matter Physics and \\ Institute of Physics, Chinese Academy of Sciences, Beijing 100190, China}

\author{Jie Wang\linkedsup{fn:jie}{$\ddagger$}}
\affiliation{International Center for Quantum Materials, Peking University, Beijing 100871, China}
\affiliation{Beijing Key Laboratory of Quantum Devices, Peking University, Beijing, 100871, China}

\author{Fu-Chun Zhang\linkedsup{fn:zhang}{$\S$}}
\affiliation{Kavli Institute for Theoretical Sciences, University of Chinese Academy of Sciences, Beijing, 100190, China}
\affiliation{School of Physical Science and Technology, ShanghaiTech University, Shanghai 201210, China}

\begin{abstract}
Moir\'e transition metal dichalcogenides have served as a versatile platform for simulating Hubbard physics. Recent experiments have identified robust superconductivity in moir\'e bilayer WSe$_2$ for certain twist angles. Here, we propose the gossamer nature of the superconductivity recently discovered at half-filling and zero displacement field in twisted WSe$_2$.
By mapping the moir\'e continuum system to an effective extended single-orbital Hubbard model on the triangular lattice, we employ renormalized mean-field theory to investigate the strong-coupling phase diagram. 
We find that a moderate Coulomb repulsion partially suppresses charge fluctuations while preserving a finite density of mobile doublons and holes. In this regime, the interplay between extended kinetic hoppings and antiferromagnetic superexchange stabilizes a chiral $d+id$ superconducting phase.
Our results naturally account for the twist-angle-dependent evolution from a Mott insulator to a superconductor and eventually to a correlated metal. Furthermore, the model demonstrates that this half-filled pairing state vanishes rapidly upon density doping, consistent with experimental observations.
\end{abstract}

\maketitle

Moir\'e materials have emerged as versatile platforms for realizing exotic many-body quantum phases, owing to their enhanced interaction effects and exceptional tunability~\cite{MoireSimulatorReview21,Mak2022,Review_Mak_Shan_HubbardMoire_26,YaoWangReview25,YazdaniReview24,Review_FCI_moire26}. 
Among them, moir\'e transition metal dichalcogenides (TMDs) serve as ideal solid-state simulators for Hubbard physics~\cite{Wu2018_Theory, Tong2017_Moire_WangYao, Pan_Wu_2020,Kuhlenkamp_PRX_2024}, hosting a rich variety of correlated phenomena including generalized Wigner crystals~\cite{Wang2020_Mott, Tang2020_Wigner, Regan2020_Mott, Xu2020_Wigner, Huang2021_Fractional}, tunable metal-insulator transitions~\cite{Ghiotto2021_MIT, Li2021_MIT, Zang2021_DMFT_Millis,Ryee2026}, Kondo lattices~\cite{Zhao2023_Kondo, Han2026,Coleman2023_MoireKondo}, and fractional Chern insulators~\cite{Bernevig_FCI11,Review_FCI_moire26, TingxinLiSC2026,LIU2024515}.

Recent experiments on twisted WSe$_2$ (tWSe$_2$) have unveiled a rich, highly tunable phase diagram controlled by the carrier density $\nu$ and an external displacement field. By dynamically tuning the correlation strength, these studies have established a comprehensive phase diagram encompassing Mott insulators, correlated metals~\cite{Wang2020_Mott,Ghiotto2021_MIT}, and most recently, robust superconducting (SC) states~\cite{Xia2025,Xia2026,Guo2025,Guo2026}. 
Notably, at half-band filling ($\nu=1$) without an electric field, the twist angle $\theta$ effectively tunes the electronic correlation, driving a remarkable sequence of transitions~\cite{Xia2025}: from a Mott insulator ($\theta \approx 2.5^\circ$), to a SC state ($\theta \approx 3.6^\circ$), and ultimately to a metal state ($\theta \approx 4.5^\circ$). The emergence of superconductivity exactly at a filling of one electron per moir\'e unit cell strongly indicates a Mott-driven mechanism~\cite{Anderson2004,Lee2006,Powell2011}.

\begin{figure}[t]
\includegraphics[width=1\linewidth]{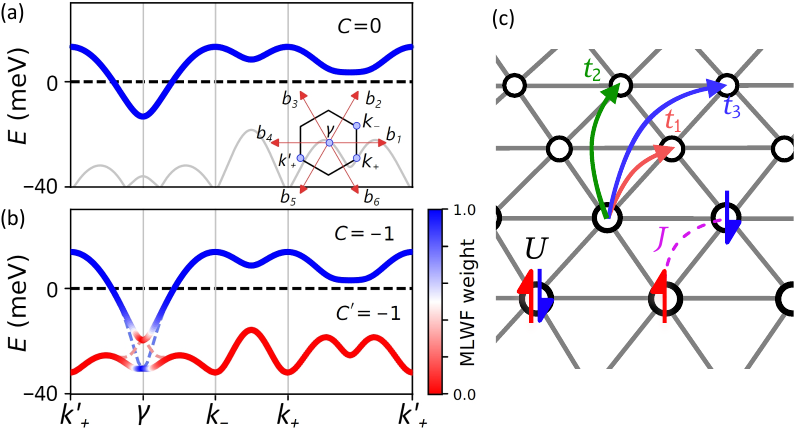} 
\caption{\textbf{Moir\'e band structures and the triangular-lattice $t_1$-$t_2$-$t_3$ Hubbard model.} 
(a) and (b) Low-energy band structure of tWSe$_2$ at $3.6^\circ$ obtained from the two sets of commonly used continuum model parameters (see Table~\ref{tab:parameters}). 
Due to uncertainties inherent in the \emph{ab initio} calculations, the fitting schemes, and experimental determination of model constants, the two parameter sets yield different topologies for the top hole band. Despite this ambiguity, in both cases, the top band is predominantly contributed by maximally localized real-space orbitals on a triangular lattice. See the text for discussion. (c) Schematic of the effective moir\'e sites (circles). The colored solid arrows indicate the isotropic electron hopping amplitudes up to the 3rd nearest-neighbor (NN) bonds: $t_1$ (red), $t_2$ (green), and $t_3$ (blue). The dominant intra-orbital Coulomb repulsion $U$ is depicted by the opposite spins (red and blue arrows) occupying the same site. The dashed line denotes the antiferromagnetic superexchange $J$ between 1st NN bonds. The inset in (a) shows the first Brillouin zone, high symmetry momentum points and the reciprocal lattice vectors.} \label{fig:fig1}
\end{figure}

Several theoretical scenarios, including topological quantum fluctuations~\cite{Si_SC_24}, extended attractions~\cite{Tuo2025_Yao_tWSe2}, and fluctuation-mediated pairings~\cite{Wu_MoireTopSC_25, Millis_MoireTopSC_24,Christos_2025,Chen_Arxiv_2025, Kuhlenkamp_Arxiv_2025,Wu2023_PRL,Fischer2025}, have been proposed to understand the pairing mechanism in tWSe$_2$. However, the puzzle of how macroscopic phase coherence develops from a half-filled Mott state without an external electric field remains unresolved.

The low-energy electronic structure of tWSe$_2$ is typically governed by three uppermost moir\'e valence bands. Depending on the system parameters, these bands exhibit either trivial (Chern number $C=0$) or nontrivial ($C=\pm 1$) topology~\cite{Wu_Macdonald_2019, Pan_Wu_2020, Devakul2021-zu,Yu2020NSR}.
Nevertheless, at the twist angles where correlated phases are most pronounced~\cite{Review_Mak_Shan_HubbardMoire_26, Xia2026}, the low-energy physics can be effectively captured by a single-orbital model, regardless of the band topology~\cite{Wu_Macdonald_2019, Pan_Wu_2020, Devakul2021-zu, Xie2025}. This crucial property allows us to reliably map the continuum system onto an effective single-orbital extended Hubbard model on a triangular lattice (see Fig.~\ref{fig:fig1}), to thoroughly examine the correlation-driven phenomena.

In this Letter, we theoretically investigate the effective triangular-lattice model and propose that the intermediate phase in tWSe$_2$ at half filling is a genuine realization of \emph{gossamer superconductivity}~\cite{Laughlin2006, Bernevig2003}. 
Intimately connected to the physics of doped Mott insulators and high-$T_c$ cuprates~\cite{Zhang2003, Coleman2003,Gan2005prb}, the gossamer SC state sustains finite doublons and holes even at half-filling, providing the requisite charge itinerancy for global phase coherence. 
By using renormalized mean-field theory~\cite{Zhang_RMFT_1988}, we show that the extended kinetic hoppings, e.g., $t_2$ and/or $t_3$, are essential for stabilizing a chiral $d+id$ gossamer SC state driven by the superexchange interaction.
Similar to the pressure-driven transitions in organic superconductors~\cite{Gan2005,Powell2011}, our ground-state phase diagram qualitatively accounts for the twist-angle-dependent evolution observed in experiments: transitioning from a Mott insulator, through the gossamer SC state, to a correlated metal. 
Furthermore, because this gossamer state inherently sustains charge itinerancy at half-filling, it is found to be highly sensitive to external density doping. The rapid suppression of pairing upon slight doping naturally explains the remarkably narrow SC dome observed experimentally.

\emph{Band structure of tWSe$_2$.---}
We start by describing the band structure of tWSe$_2$. The strong spin-orbit coupling in WSe$_2$  locks the spin of the active hole band to its valley degree of freedom, and the spatial twisting further hybridizes the layer components. The long-wavelength effective description of the twisted homobilayer WSe$_2$ is captured by the continuum model. Near the $\bm K_+$ valley, in the basis $\Psi^\dagger_{\uparrow}=(\psi^\dagger_{b\uparrow},\psi^\dagger_{t\uparrow})$, the continuum model is given by~\cite{Wu_Macdonald_2019}:
\begin{equation*}
H_{\uparrow}= \left(\begin{array}{cc}
-\dfrac{\hbar^{2}({\bm k}-\bm{\kappa}_{+})^{2}}{2m^*}
+\Delta_{+}(\bm r) & \Delta_T(\bm r) \\
\Delta_T^{\dagger}(\bm r) & -\dfrac{\hbar^{2}({\bm k}-\bm{\kappa}_{-})^{2}}{2m^*} +\Delta_{-}(\bm r)
\end{array}\right)
\end{equation*}
and the spin-down Hamiltonian at the $\bm K_-$ valley is obtained by time-reversal transformation. In the above, $m^*$ is the effective mass, $\bm{\kappa}_\pm$ denote the momentum offsets of the two layers due to twisting effects. The continuum model is determined by the symmetry of the twisted TMD system, which includes a three-fold rotation $C_{3z}$ around the $\hat{z}$ axis and a two-fold rotation $C_{2y}$ around the $\hat{y}$ axis. $\Delta_{\pm}(\bm r) = 2V \sum_{i=1,3,5} \cos\left(\bm b_i\cdot\bm r \pm \psi\right)$ and $\Delta_T(\bm r) = w \left(1+e^{-i\bm b_2\cdot\bm r} + e^{-i\bm b_3\cdot\bm r}\right)$ are respectively the intra- and interlayer moir\'e potential. 
The $\bm b_i$ are the six reciprocal lattice vectors of the triangular moir\'e lattice [see the inset of Fig.~\ref{fig:fig1}(a) for an illustration].

The parameters of the continuum model $(\psi, w, V)$ are determined by fitting to first-principles calculations and experimental data.
However, due to ambiguities inherent in the \emph{ab initio} calculations and the fitting schemes, the resulting band structure and band topology are subject to a certain degree of uncertainty~\cite{Wu_Macdonald_2019,Devakul2021-zu,Xiao_Polarization_NC24,Zhang2025,zhang2024}. Fig.~\ref{fig:fig1} (a) and (b) illustrate the top two active bands of tWSe$_2$ at 3.6$^\circ$, obtained from two commonly used sets of continuum model parameters summarized in Table~\ref{tab:parameters}. Owing to uncertainties in the determination of the model parameters, even the predicted topology of the active bands becomes ambiguous. Nevertheless, we argue below that such ambiguity in continuum-model parameterization
does not invalidate the use of a single-band Hubbard model for the low-energy effective description of the system.

\begin{table}[t]
\begin{tabular}{ c || c | c | c || c || c}
    \hline
    \hline
    & $\psi$ & $w$ & $V$ & $C$ & Reference \\ \hline
    Fig.~\ref{fig:fig1} (a) & $49.1^\circ$ & $10~\mathrm{meV}$ & $13.6~\mathrm{meV}$ & $0$ & Ref.~\onlinecite{Zhang2025} \\
    Fig.~\ref{fig:fig1} (b) & $128^\circ$ & $18~\mathrm{meV}$ & $9~\mathrm{meV}$ & $-1$ & Ref.~\onlinecite{Devakul2021-zu}   \\
    \hline
    \hline
\end{tabular}
\caption{Continuum model parameters, top band Chern number and corresponding references for tWSe$_2$ at $3.6^\circ$, used in Fig.~\ref{fig:fig1} (a) and (b).} \label{tab:parameters}
\end{table}

For the non-topological set of parameters [Fig.~\ref{fig:fig1}(a)], the top hole band is topologically trivial and can be straightforwardly Wannierized~\cite{Marzari_1997,Souza_2001,pizzi2020wannier90}. 
Conversely, for the topological set of parameters [Fig.~\ref{fig:fig1}(b)], the top band is topological. Nevertheless, a partial Wannierization scheme can be utilized to extract a maximally localized orbital alongside a delocalized counterpart, whose weight is sharply concentrated near the $\gamma$ point~\cite{Xie2025}.
In both the trivial and topological regimes, the localized orbital resides on a triangular lattice and dominates the real-space density of the active hole band across the vast majority of the Brillouin zone. Since electrons on this localized orbital are subject to strong Coulomb repulsion, it primarily drives the local correlation physics, including Mott localization and short-range superexchange. 
Based on these considerations, we propose an effective single-orbital extended Hubbard model on the triangular lattice for the half-filled tWSe$_2$ where the superconductivity is observed.

\emph{Effective interacting model.---}
To capture the low-energy physics and the strongly-correlated nature of the moir\'e valence band, we map the continuum system onto an effective single-orbital extended Hubbard model on the triangular lattice. The full Hamiltonian, defined on the lattice shown in Fig.~\ref{fig:fig1}(c), reads $\mathcal{H} = \mathcal{H}_0 + \mathcal{H}_U + \mathcal{H}_J$, where
\begin{equation}
\begin{split}
    \mathcal{H}_0 &= -\sum_{n=1,2,3}\sum_{\langle i j \rangle_n, \sigma} t_n\left(c^\dagger_{i, \sigma} c_{j, \sigma} + \text{h.c.}\right), \\
    \mathcal{H}_U &= U \sum_{i} n_{i\uparrow} n_{i\downarrow}, \quad 
    \mathcal{H}_J = J\sum_{\langle i j \rangle_1}{\bf S}_i\cdot {\bf S}_j.
\end{split}\label{eq:ham}
\end{equation}
Here, $c^\dagger_{i, \sigma}$ creates a fermion at the moir\'e site $i$ with spin-valley index $\sigma \in \{\uparrow, \downarrow\}$. The kinetic energy $\mathcal{H}_0$ is parameterized by the $t_1$-$t_2$-$t_3$ hoppings on the $n$-th nearest-neighbor (NN) bonds $\langle i j \rangle_n$. The strong correlation is governed by the on-site Hubbard repulsion $\mathcal{H}_U$ acting on the local density $n_{i\sigma} = c^\dagger_{i, \sigma} c_{i, \sigma}$. The parameters $t_n$ are extracted from the aforementioned continuum model calculations for tWSe$_2$; see details in the Supplemental Material~\cite{SM}.

In addition to the standard $\mathcal{H}_0$ and $\mathcal{H}_U$, we explicitly introduce an antiferromagnetic superexchange $\mathcal{H}_J$ on the 1st NN bonds, which couples the local spins ${\bf S}_i = \frac{1}{2}\sum_{\alpha\beta} c^\dagger_{i,\alpha} \boldsymbol{\sigma}_{\alpha\beta} c_{i,\beta}$. This term is crucial for properly capturing the emergent interactions in the large-$U$ limit, and is appropriate in the variational approach studied here~\cite{Zhang2003}. In the weak-coupling regime (small $U$), a small $J$ has a negligible impact on the itinerant normal state. Nevertheless, as $U$ increases towards the strong-coupling limit, charge fluctuations are heavily penalized, and $\mathcal{H}_J$ naturally emerges as the dominant force dictating the low-energy physics. Therefore, the explicit inclusion of $J$ in Eq.~\eqref{eq:ham} provides a unified framework to systematically investigate strong-coupling phenomena, such as gossamer SC states,  by simply tuning $U$.

\begin{figure}[!b]
    \centering
    \includegraphics[width=0.95\linewidth]{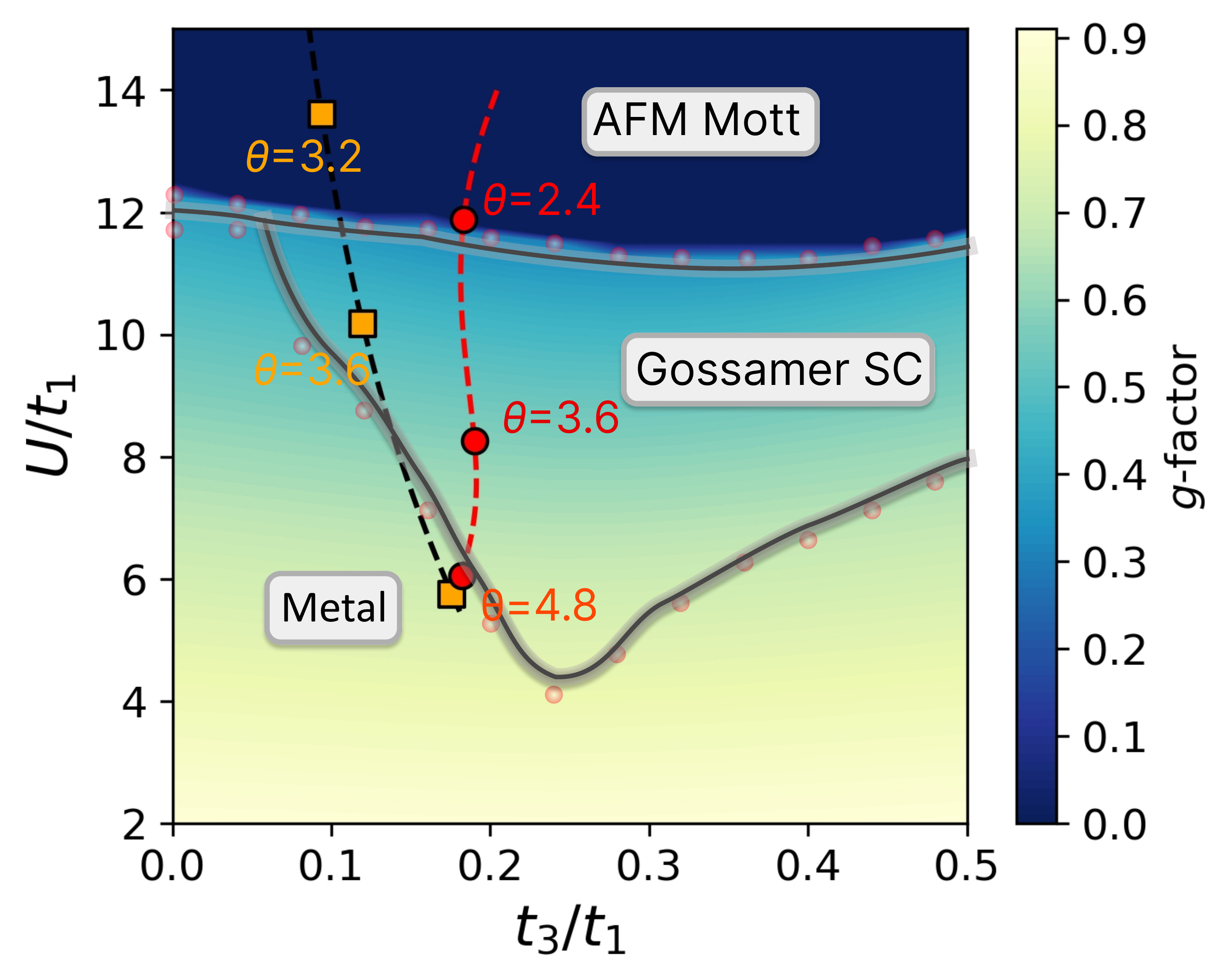}
    \caption{\textbf{Half-filling ($\nu=1$) ground-state phase diagram in parameter space $U/t_1$ and $t_3/t_1$, for Hamiltonian in Eq.~\eqref{eq:ham} with $t_2=0$ and $J/t_1=0.25$.} The background color map indicates the optimized translationally-invariant Gutzwiller factor $g$. Solid black lines delineate the phase boundaries between the correlated metal, the gossamer superconductor (SC), and the $120^\circ$ AFM Mott insulator.
    The black and red dashed lines represent the theoretical trajectories of the effective parameters for twisted bilayer WSe$_2$ as functions of the twist angle $\theta$ (in degrees), corresponding to the trivial and topological bands in Figs.~\ref{fig:fig1}(a) and (b), respectively.
    Here, the interaction strength scales as $U = e^2/(4\pi \epsilon_0 \epsilon_r a_M)$, where $\epsilon_r$ denotes the effective dielectric constant and is taken to be $10$.
    }
    \label{fig:phasediagram}
\end{figure}

Throughout this work, we set $t_1=1$ as the energy unit and fix $J=0.25$, systematically tuning $t_2$, $t_3<t_1$, and $U$ to explore the ground-state phase diagram.

\emph{RMFT theory.---}
To map out the strongly-correlated phases in Eq.~\eqref{eq:ham}, we employ the renormalized mean-field theory (RMFT)~\cite{Zhang_RMFT_1988}. Our starting point is the Gutzwiller-projected wavefunction $|\Psi_G(\{\alpha\})\rangle = P_G |\Psi_{\text{MF}}(\{\alpha\})\rangle$, which captures the core physics through two synergistic components. The operator $P_G = \prod_{i} [1 - (1-g_{i}) n_{i\uparrow} n_{i\downarrow}]$ introduces a local variational weight $g_{i} \in [0,1]$. Naturally, it suppresses on-site double occupancies to minimize the dominant Hubbard repulsion $\mathcal{H}_U$. Under this local projection, the unprojected state $|\Psi_{\text{MF}}(\{\alpha\})\rangle$ acts as a versatile fermionic vacuum to characterize different underlying physical phases, where the parameters  $\{\alpha\}$ encode different physical states, such as superconductors, spin liquids, or magnetic orders. The lowest-energy ground state is then obtained by optimizing $g_i$ and $\{\alpha\}$ to minimize the variational energy $E_G = \langle \mathcal{H}\rangle_G$, with $\langle \cdots \rangle_G\equiv{}\langle \Psi_G | \cdots | \Psi_G \rangle / \langle \Psi_G | \Psi_G \rangle$.

However, analytically evaluating the energy expectation value $E_G$ is generally intractable. Here, we utilize the Gutzwiller approximation, which approximates the non-local correlation effects in $P_G$ as simple statistical counting. In this framework, the physical double occupancy $d_i = \langle n_{i\uparrow}n_{i\downarrow} \rangle_G$ at site $i$ is estimated at the tree level as \[d_i=g_i^2 d_{0}/[1 - (1-g_i^2)d_{0}],\] where $g_{i}$ is the variational Gutzwiller factor to be optimized and $d_0$ is the bare double occupancy with respect to $|\Psi_{\text{MF}}(\{\alpha\})\rangle$. Note that the mean-field ansatz we use here ensures the translational invariance of $d_0$. The suppression of $d_i$ inevitably redistributes the statistical probabilities of the other three local configurations: holes, spin-up, and spin-down. Since electron hopping and spin exchange depend strictly on the availability of these specific configurations, their effective amplitudes are heavily modified. The approximation captures this by dressing the bare kinetic and exchange operators with renormalization factors that reflect this altered local phase space, yielding the effective Hamiltonians $\tilde{\mathcal{H}}_0$ and $\tilde{\mathcal{H}}_J$. Consequently, the variational energy can be evaluated simply with respect to the unprojected state as $E_G \approx \langle \Psi_{\text{MF}} | \tilde{\mathcal{H}}_0 + \tilde{\mathcal{H}}_J | \Psi_{\text{MF}} \rangle + U \sum_i d_i$. The explicit expressions for $\tilde{\mathcal{H}}_{0}$ and $\tilde{\mathcal{H}}_{J}$, which can be analytically expressed in terms of $g_i$ and the local densities of $|\Psi_{\text{MF}}\rangle$~\cite{SM}. Within the present RMFT, $U$ controls the suppression of charge fluctuations, whereas $J$ provides the explicit spin-singlet pairing channel. Here, we keep $J/t_1$ fixed to isolate the evolution with $U/t_1$; see results obtained with $J=4t_1^2/U$ in Supplemental Material~\cite{SM}.

The unprojected vacuum $|\Psi_{\text{MF}}\rangle$ is constructed as the ground state of the following trial mean-field Hamiltonian:
\begin{equation}
\begin{split}
H_{\text{MF}} = &\sum_{i,j} \left[ -\chi_{ij} \sum_\sigma c^\dagger_{i\sigma} c_{j\sigma} + \left( \Delta_{ij} c^\dagger_{i\uparrow} c^\dagger_{j\downarrow}+ \text{h.c.} \right) \right] \\
& + \sum_i \left( -\mu n_{i} + \mathbf{m}_i \cdot \mathbf{S}_i\right).    
\end{split}
\end{equation}
The parameters $\{\chi_{ij}, \Delta_{ij}, \mathbf{m}_i, \mu\}$ are optimized to minimize the variational energy $E_G$. 
Specifically, the hoppings $\chi_{ij}$ are retained up to the 3rd nearest-neighbor (NN) bonds. The pairing amplitudes $\Delta_{ij}$, naturally motivated by $H_J$, are restricted to the first NN bonds. The local field $\mathbf{m}_i$ characterizes possible magnetic orders, and a chemical potential $\mu$ is tuned to control the electron filling. 

To study the ground state from weak to strong coupling limits, we propose three competing ans\"{a}tze: 
(i) In the weak-coupling limit, the essential physics is captured by a correlated metal retaining solely uniform hoppings ($\Delta_{ij}=0$ and $\mathbf{m}_i=0$), where interactions primarily renormalize the 2nd and 3rd hopping amplitudes, $\chi_2$ and $\chi_3$, respectively.
(ii) A SC state with finite singlet pairing $\Delta_{ij}$ emerges from the correlated metal as $U$ increases. We propose extended $s$-wave, nematic $d$-wave, and chiral $d+id$-wave ans\"{a}tze, where the $d$-wave channels belong to the 2D $E_2$ irreducible representation of the triangular lattice.
(iii)  In the large-$U$ Mott regime dominated by magnetic frustration, we activate a noncollinear $120^\circ$ N\'eel order via a finite $\mathbf{m}_i$. We additionally allow a fixed staggered-flux hopping background~\cite{Zhou_Wen_2002,Lu_Triangular_2016} in the variational ansatz. The details of the mean-field ans\"{a}tze can be found in the Supplemental Material~\cite{SM}.

\emph{Theoretical phase diagram, experimental comparison and implications.---}
We focus on the half-filled case ($\nu=1$). Guided by Wannierization calculations that the leading extended hopping term is $t_3$, we first consider the $t_2=0$ case. The ground state in the $U/t_1$-$t_3/t_1$ parameter space is then identified by comparing the optimized energies of the three proposed ans\"{a}tze.

The half-filled ground-state phase diagram is illustrated in Fig.~\ref{fig:phasediagram}. We find that the optimized Gutzwiller factor $g_i=g$ preserves the translational invariance~\footnote{Note that the 120$^\circ$ AFM Mott state breaks translational symmetry in the spin sector but preserves it in charge sector, which also leads to uniform $g_i$.}. The background colormap represents the optimized Gutzwiller factor $g$, which acts as a continuous barometer for charge itinerancy---ranging from a weakly correlated metal ($g \rightarrow 1$) to a Mott insulator ($g = 0$). For a small extended hopping ($t_3/t_1 \lesssim 0.1$), increasing $U$ simply drives a direct metal-insulator transition. In the strong-coupling Mott regime, the on-site charge fluctuations are completely frozen. The low-energy physics is then governed by localized spins. Our energetic optimization reveals that this insulating state is stabilized by a coplanar $120^\circ$ N\'eel magnetic order with a fixed staggered-flux hopping background.

\begin{figure}[!t]
\centering
\includegraphics[width=1.\linewidth]{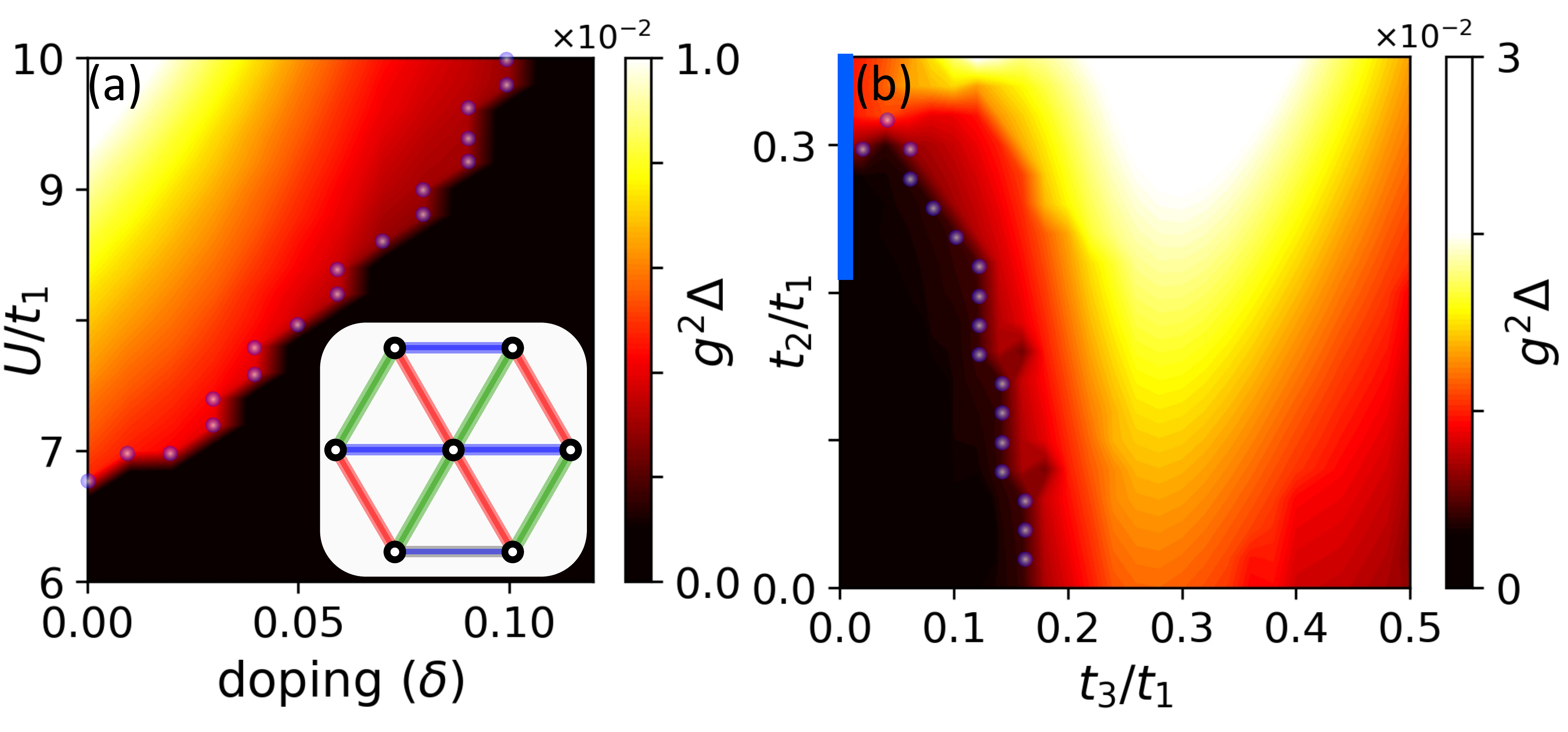}
\caption{ (a) Evolution of the gossamer SC in the $U/t_1$ versus doping ($\delta$) plane. Here we set $t_2/t_1=0.02$, $t_3/t_1=0.18$, and $J/t_1=0.25$. Inset: Schematic illustration of the optimized chiral $d+id$-wave pairing symmetry on the triangular lattice, where the blue, green, and red bonds denote the pairing form factors $1$, $e^{i2\pi/3}$, and $e^{i4\pi/3}$, respectively. (b) The half-filling projected pairing amplitude $g^2\Delta$ in the $t_2$-$t_3$ parameter space, with $U/t_1 = 8$ and $J/t_1 = 0.25$. The blue line marks the purely $t_2$-driven SC phase recently identified by DMRG studies in Refs.~\cite{Huang2023, Zhu2023}. }
\label{fig:pairing}
\end{figure}

Notably, beyond a threshold of $t_3$, the interplay between the extended kinetic energy and moderate $U$ preempts this direct Mott transition, giving way to an intermediate SC phase. This state is characterized by a partially suppressed charge itinerancy ($g \sim 0.5$--$0.7$) with a robust $d+id$-wave pairing; see the pairing symmetry in Fig.~\ref{fig:pairing}(a). We emphasize that in the framework of RMFT, the unprojected amplitude $\Delta$ captures the short-range singlet pairing correlations driven by the superexchange $J$, while the physical SC order parameter is strongly renormalized to $g^2\Delta$~\cite{Zhang2003,Edegger2007,SM}. This renormalization accounts for the partial projection of doubly occupied states. Unlike a strict Mott insulator where on-site charge fluctuations are completely forbidden, the system maintains a finite mobile carrier density at half-filling. The resulting phase---sustaining robust microscopic pairing alongside a heavily suppressed superfluid density---is identified as a gossamer SC~\cite{Laughlin2006,Zhang2003}.

We propose that this theoretical phase diagram naturally captures the recent experimental observation of SC in twisted bilayer WSe$_2$. To show this, we overlay the trajectory of the effective model parameters tuned by twist angle $\theta$; see the dashed line in Fig.~\ref{fig:phasediagram}. At small twist angles ($\theta \lesssim 3^\circ$), the extreme flatness of the moir\'e bands yields a dominant $U/t_1$, placing the system deep inside the Mott insulating regime. As $\theta$ increases, the moir\'e bandwidth expands, which simultaneously reduces $U/t_1$ and enhances $t_3/t_1$. Following this physical trajectory, the system is driven out of the Mott state directly into the gossamer SC phase,  in qualitative agreement with the experimental onset of superconductivity around $\theta\approx3.6^\circ$. 
Upon further increasing the twist angle, the effective $U/t_1$ continues to drop. The weakened electronic correlations eventually fail to sustain a finite pairing, leaving behind a correlated metal. 
We note that the experimental trajectories in Fig.~\ref{fig:phasediagram} should be viewed as qualitative estimates whereas the intermediate gossamer-SC regime is robust against uncertainties in the effective parameters such as dielectric screening~\cite{SM}. The displacement-field-induced suppression of superconductivity in favor of the insulating state is discussed in the Supplemental Material~\cite{SM}.

Recent experiments on tWSe$_2$ observe an abrupt destruction of SC upon slight doping away from half-filling~\cite{Xia2026}. To understand this, we compute the physical pairing amplitude $g^2\Delta$ versus Hubbard $U$ and doping $\delta$; see Fig.~\ref{fig:pairing}(a). The phase diagram reveals that the SC is optimal at $\delta=0$. Crucially, in the moderately correlated regime (near the lower phase boundary), this gossamer state is remarkably fragile, vanishing rapidly upon slight doping. In contrast, at larger $U/t_1$, the SC dome broadens and survives up to higher hole/electron concentrations. 

This doping dependence reflects the distinction between the gossamer SC and the general high-$T_c$ SC in doped Mott systems. At large $U/t_1$, finite doping is therefore essential to activate the charge degrees of freedom in the rigid Mott state, naturally broadening the SC dome. Conversely, the partially projected gossamer state ($g>0$) inherently supports charge fluctuations at half-filling. Upon doping, the enhanced bare kinetic energy rapidly outweighs the superexchange $J$, suppressing the SC pairing. This mechanism naturally explains the highly restricted SC phase space observed experimentally.

To gain insight into the $t_3$-driven SC, we note recent DMRG studies demonstrating that $t_2$ solely can induce SC in the triangular-lattice $t$-$J$ model~\cite{Huang2023, Zhu2023}. This suggests that $t_3$ and $t_2$ may play analogous roles in promoting SC. To explicitly verify this analogy and explore their interplay, we map out the half-filled phase diagram in the $t_2$-$t_3$ parameter space. As demonstrated in Fig.~\ref{fig:pairing}(b), either a finite $t_2$ or $t_3$ alone is sufficient to induce the gossamer SC. More importantly, their simultaneous presence can enhance the pairing strength, leading to a more robust SC regime. Across this entire regime, the optimized state consistently favors a chiral $d+id$-wave pairing symmetry, which qualitatively agrees with the $d$-wave nature extracted from DMRG~\cite{Huang2023} and dynamical mean-field theory~\cite{Belanger2022} calculations.

\emph{Discussion. ---}
In summary, we propose that a one-band gossamer superconductivity provides a consistent framework for understanding the half-filled pairing state in tWSe$_2$.
By solving the effective triangular-lattice extended Hubbard model, we showed that the 2nd and/or 3rd NN  hoppings ($t_2, t_3$), together with AFM superexchange, stabilize this gossamer SC phase. 
The proposed gossamer picture, characterized by partially unfrozen charge fluctuations, naturally accounts for both the twist-angle-driven Mott-to-SC transition and the extreme fragility of the pairing against density doping.

The predicted $d+id$ pairing state breaks time-reversal symmetry, which may be probed in optical measurements such as Kerr rotation~\cite{Kapitulnik2009Kerr,Nandkishore2014Kerr}. Incorporating superconducting phase fluctuations beyond RMFT is an important direction for future study. Nevertheless, we provide rough estimates of the $T_c$ and coherence length within the RMFT framework in the Supplemental Material~\cite{SM}.
Conceptually, our minimal proposal establishes moir\'e TMDs as highly controllable Hubbard simulators, paving the way for future explorations of elusive Mott-driven phenomena such as the pseudogap and quantum phase transitions. 

\begin{acknowledgements}
\emph{Acknowledgements.---} We acknowledge Kin Fai Mak, Zhongdong Han, Tingxin Li and Fang Xie for useful discussions.
H.-K. J. acknowledge the support by the National Natural Science Foundation of China (NSFC-12504180), and the start-up funding from ShanghaiTech University. G. J. and J. W. acknowledge the support by the National Natural Science Foundation of China (Grant No. 12674196),  the Quantum Science and Technology-National Science and Technology Major Project (Grant No. 2025ZD0300500) and the Fundamental Research Funds for the Central Universities, Peking University. F.-C. Z. is partly supported by Ministry of Science and Technology (Grant No. 2022 YFA1403900), Innovation Program for Quantum Science and Technology (Grant No. 2021ZD0302500), and The Basic Research Program of the Chinese Academy of Sciences Based on Major Scientific Infrastructures (Grant No. JHKYPT-2021-08), and NSFC grant 12574150.  
\end{acknowledgements}

\vspace{0.5cm}
\centerline{\rule{2.7cm}{0.4pt}}
\vspace{0.3em}
\begin{myfootnotes}
\myfootnote{$^{*}$}{fn:common}{These authors contributed equally.}
\myfootnote{$^{\dagger}$}{fn:jin}{ \href{mailto:jinhk@shanghaitech.edu.cn}{jinhk@shanghaitech.edu.cn}}
\myfootnote{$^{\ddagger}$}{fn:jie}{\href{mailto:jiewang.phy@pku.edu.cn}{jiewang.phy@pku.edu.cn}}
\myfootnote{$^{\S}$}{fn:zhang}{\href{mailto:fuchun@ucas.ac.cn}{fuchun@ucas.ac.cn}}
\end{myfootnotes}

\bibliography{main}

\clearpage
\newpage

\begin{widetext}

\begin{center}

{\large\bf Supplemental Material for``Gossamer Superconductivity in Moir\'e WSe$_2$ Bilayer"}
\vspace{0.5cm}
\\
Hui-Ke Jin$^{*, \dagger,1}$, Guangyue Ji$^{*,2,3}$, Zhan Wang$^{4}$,
Jie Wang$^{\ddagger,2,3}$, and Fu-Chun Zhang $^{\S
,5,1}$
\vspace{0.3cm}

{\em $^1$School of Physical Science and Technology, ShanghaiTech University, Shanghai 201210, China}

{\em $^2$International Center for Quantum Materials, Peking University, Beijing 100871, China}

{\em $^3$Beijing Key Laboratory of Quantum Devices, Peking University, Beijing, 100871, China}

{\em $^4$Beijing National Laboratory for Condensed Matter Physics and \\ Institute of Physics, Chinese Academy of Sciences, Beijing 100190, China}

{\em $^5$Kavli Institute for Theoretical Sciences, University of Chinese Academy of Sciences, Beijing, 100190, China}
\vspace{0.5cm}

\end{center}


In this Supplemental Material, we provide detailed theoretical derivations and additional numerical results supporting the main text. Section A introduces the continuum model and partial-Wannierization procedure, examines the validity of the effective single-orbital description and the role of quantum geometry, and analyzes the effects of an out-of-plane displacement field. Section B presents the formulation of the renormalized mean-field theory (RMFT) within the Gutzwiller approximation, while Section C defines the normal, superconducting, and magnetic variational ans\"atze used in our calculations. Section D provides further RMFT results clarifying the roles of the superexchange interaction $J$, the extended hoppings $t_2$ and $t_3$, doping, and dielectric-screening uncertainty in the stability of the gossamer superconducting state. Finally, Section E discusses the limitations associated with phase fluctuations and provides order-of-magnitude estimates of the superconducting transition temperature and coherence length.

\section{A. Continuum model and Wannierization}
\subsection{A1. The continuum model of moir\'e TMD}

The continuum kinetic Hamiltonian for $t\mathrm{WSe}_{2}$ near the $K_+$ valley is given by
\begin{equation}
H_{0,\uparrow}=
\begin{bmatrix}
-\frac{\hbar^{2}(-i \bm{\nabla}-\bm{\kappa}_{+})^{2}}{2m^*}
+\Delta_{+}(\bm r)
&
\Delta_T(\bm r)
\\
\Delta_T^{\dagger}(\bm r)
&
-\frac{\hbar^{2}(-i \bm{\nabla}-\bm{\kappa}_{-})^{2}}{2m^*}
+\Delta_{-}(\bm r)
\end{bmatrix},\label{eq:continuum_model}
\end{equation}
where the $2\times2$ matrix acts in the layer pseudospin space 
$\Psi^\dagger=(\psi_b^\dagger,\psi_t^\dagger)$. In the above, $\Delta_{\pm}(\bm r) = 2V \sum_{i=1,3,5} \cos(\bm b_i \cdot \bm r \pm \psi)$ is the intralayer moir\'e potential. The $\Delta_T(\bm r) = w \left(1+e^{-i\bm b_2\cdot\bm r}+e^{-i\bm b_3\cdot\bm r}\right)$ denotes the interlayer tunneling where $\bm{b}_{i}$ are six reciprocal lattice vectors.
The form of the continuum model is fixed (up to layer-dependent phases) by symmetries which include~\cite{Christos_2025}:
\begin{itemize}
\item \textbf{Threefold rotation $C_{3z}$.}
Under a $2\pi/3$ rotation around the $z$ axis, the model is invariant,
\begin{equation}
C_{3z}:
\begin{pmatrix}
x\\y
\end{pmatrix}
\rightarrow
\begin{pmatrix}
\cos\frac{2\pi}{3} & -\sin\frac{2\pi}{3}\\
\sin\frac{2\pi}{3} & \cos\frac{2\pi}{3}
\end{pmatrix}
\begin{pmatrix}
x\\y
\end{pmatrix}.
\end{equation}
\item \textbf{Time-reversal symmetry.}
The two valleys are related by the anti-unitary operator $\Theta = i\eta_y \mathcal{K}$ where $\eta_j$ are Pauli matrices acting in valley space and $\mathcal{K}$ denotes complex conjugation.
\item \textbf{Mirror reflection $m_x$.}
Reflection with respect to the $xz$ plane acts as $m_x: \Psi(x,y) \rightarrow i\eta_y \Psi(x,-y)$.
\item \textbf{Intravalley inversion symmetry.}
Unitary symmetry $\mathcal{I}:\Psi(\mathbf r)\rightarrow \ell_x\,\Psi(-\mathbf r)$, where $\ell_j$ are Pauli matrices acting in layer space. This is a symmetry of the moir\'e TMD when the displacement field is absent.
\end{itemize}

$C_{2y}$ in the main text can be written as the product of the mirror reflection $m_x$ and the intravalley inversion symmetry $\mathcal I$.

\subsection{A2. Partial Wannierization of the Top Moir\'e Bands} \label{sec:partial_wannierization}

In the symmetry analysis and in the construction of the input matrices for partial Wannierization, it is convenient to eliminate the layer-dependent momentum shifts by a gauge transformation $U(\bm r)=
\mathrm{diag}
\left(
e^{i\bm\kappa_{+}\cdot\bm r},
e^{i\bm\kappa_{-}\cdot\bm r}
\right)$. Under this transformation the Hamiltonian becomes
\begin{equation}
    H'_{0,\uparrow} = \left[ -\frac{\hbar^2\nabla^2}{2m^*} + \Delta_0(\bm r) \right]\sigma_0 + \bm{\Delta}(\bm r)\cdot\bm{\sigma},
\end{equation}
where $\sigma_j$ are Pauli matrices acting in layer space. In the above, $\Delta_x = \mathrm{Re}\,\Delta_T'$, $\Delta_y = -\mathrm{Im}\,\Delta_T'$, $\Delta_z = \frac{\Delta_+ - \Delta_-}{2}$ and $\Delta_0 = \frac{\Delta_+ + \Delta_-}{2}$, where
the transformed interlayer tunneling is,
\begin{equation}
    \Delta_T'(\bm r) = \Delta_T(\bm r) e^{i(\bm\kappa_{-}-\bm\kappa_{+})\cdot\bm r} = w \sum_{j=1}^{3} e^{i\bm q_j\cdot\bm r},
\end{equation}
where the three wave-vectors satisfy $\bm q_1=\bm\kappa_{-}-\bm\kappa_{+}$, $\bm q_2=R\left(\frac{2\pi}{3}\right)\bm q_1$ and $\bm q_3=R\left(\frac{4\pi}{3}\right)\bm q_1$. The Hamiltonian in the opposite valley is obtained from time-reversal symmetry $H'_{0,\downarrow} = (H'_{0,\uparrow})^*$.

For the parameter regime shown in Fig. 1(b) in the main text, the top two moir\'e valence bands carry a nonzero total Chern number and therefore do not admit two exponentially localized Wannier orbitals simultaneously. In this case, we construct a single exponentially localized Wannier orbital by \emph{partial Wannierization}, following the disentanglement procedure implemented in \textsc{Wannier90}.

At each moir\'e crystal momentum $\mathbf{k}$, we first consider the two-dimensional Bloch subspace spanned by the top two continuum-model eigenstates $\mathcal{H}_{\mathbf{k}}=\mathrm{span}\left\{
\ket{\Psi_{1\mathbf{k}}},\ket{\Psi_{2\mathbf{k}}}
\right\}$. We choose a trial Gaussian orbital $\ket{g_{\mathbf{k}}}$ and compute its projection onto this subspace $A_n(\mathbf{k}) = \braket{\Psi_{n\mathbf{k}}|g_{\mathbf{k}}}$ where $n=1,2$. These overlaps are written into the \texttt{.amn} file and provide the initial guess for the disentanglement direction. The overlap matrices between neighboring $\mathbf{k}$ points $M_{mn}^{(\mathbf{k},\mathbf{b})} = \braket{u_{m\mathbf{k}}|u_{n,\mathbf{k}+\mathbf{b}}}$ are written into the \texttt{.mmn} file. Special care is required when $\mathbf{k}+\mathbf{b}$ lies outside the first moir\'e Brillouin zone: in that case the Bloch states must be mapped back with the appropriate reciprocal-lattice embedding in order to preserve the periodic Bloch gauge. This step is essential for obtaining stable Wannier centers and spreads.

Using the input \texttt{.amn}, \texttt{.mmn}, and \texttt{.eig} files, \textsc{Wannier90} returns the rectangular disentanglement matrix $U^{\mathrm{dis}}_{n f}(\mathbf{k})$ which selects a one-dimensional smooth subspace inside $\mathcal{H}_{\mathbf{k}}$. The Bloch wave function of the partially Wannierized orbital is then
\begin{equation}
\ket{\widetilde{u}_{f\mathbf{k}}} = \sum_{n=1}^{2}
\ket{u_{n\mathbf{k}}}\,
U^{\mathrm{dis}}_{n f}(\mathbf{k}). \label{eq:udis_f}
\end{equation}
After the subsequent gauge-fixing step of \textsc{Wannier90}, this yields a single exponentially localized Wannier orbital $\ket{W_f(\mathbf{R})} = \frac{1}{\sqrt{N_k}}
\sum_{\mathbf{k}} e^{-i\mathbf{k}\cdot\mathbf{R}} \ket{\widetilde{u}_{f\mathbf{k}}}$.

Because only one localized Wannier orbital is constructed, the complementary state in the top-two-band subspace is not exponentially localized in general. Nevertheless, it can still be defined at each $\mathbf{k}$ as the orthogonal complement to $\ket{\widetilde{u}_{f\mathbf{k}}}$. Denoting this state by $\ket{\widetilde{u}_{c\mathbf{k}}}$, one obtains an effective two-orbital description in the basis $\left\{
\ket{\widetilde{u}_{f\mathbf{k}}},
\ket{\widetilde{u}_{c\mathbf{k}}}
\right\}$. The corresponding matrix elements are,
\begin{equation}
\epsilon_f(\mathbf{k}) = \bra{\widetilde{u}_{f\mathbf{k}}} H(\mathbf{k}) \ket{\widetilde{u}_{f\mathbf{k}}},\quad \epsilon_c(\mathbf{k}) = \bra{\widetilde{u}_{c\mathbf{k}}} H(\mathbf{k}) \ket{\widetilde{u}_{c\mathbf{k}}},\quad V_{\mathrm{hyb}}(\mathbf{k}) = \bra{\widetilde{u}_{f\mathbf{k}}} H(\mathbf{k}) \ket{\widetilde{u}_{c\mathbf{k}}}.
\end{equation}
Fourier transforming $\epsilon_f(\mathbf{k})$ gives the single-orbital tight-binding model reported in the main text.

In practice, we verify the success of the partial Wannierization by checking that the Wannier center remains pinned at the origin and that the gauge-invariant spread $\Omega_I$ and total spread $\Omega$ vary smoothly with twist angle. We further extract the effective hopping amplitudes $t_1$, $t_2$, and $t_3$ from the resulting \texttt{hr.dat} file using the first three shells of the triangular moir\'e lattice.

\begin{figure}[!t]
    \centering
    \includegraphics[width=0.6\linewidth]{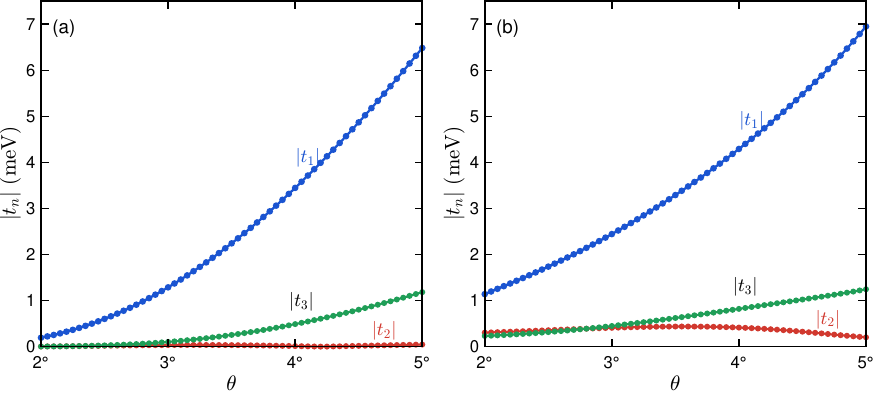}
    \caption{
    Comparison of hopping parameters extracted from full and partial Wannierization in twisted WSe$_2$.
    (a) $|t_1|$, $|t_2|$, and $|t_3|$ for the trivial band from full Wannierization.
    (b) $|t_1|$, $|t_2|$, and $|t_3|$ for the topological band from partial Wannierization.
    }
    \label{fig:hopping_two_panel}
\end{figure}

Moreover, we calculate the real-space density distributions of the constructed Wannier orbitals.
Figure~\ref{fig:wannier_density} compares the Wannier-orbital density profiles in the topologically trivial and nontrivial cases, obtained using conventional full Wannierization and the partial Wannierization procedure described above, respectively.
Both orbitals are localized around the moir\'e lattice sites $\mathcal R_M^M$ and form triangular-lattice patterns, while the orbital in the trivial band is visibly more concentrated.

\begin{figure}[t]
    \centering
    \includegraphics[width=0.6\linewidth]{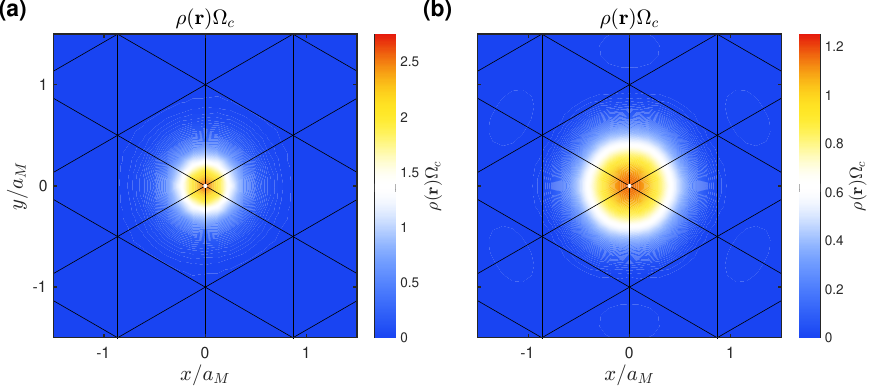}
    \caption{
    Real-space density distributions of Wannier orbitals.
    (a) Density $\rho(\mathbf r)\Omega_c$ of the fully Wannierized orbital in the topologically trivial band.
    (b) Density $\rho(\mathbf r)\Omega_c$ of the orbital obtained from partial Wannierization in the non-trivial two-band subspace.
    The coordinates are measured in units of the moir\'e lattice constant $a_M$, and the black lines denote the moir\'e lattice.
    }
    \label{fig:wannier_density}
\end{figure}

\subsection{A3. Quantum geometry and validity of the single-orbital projection}

In this subsection, we foucs on the case such that the top moir\'e valence band carries a nonzero Chern number. In this case, the topological Wannier obstruction prevents the construction of two exponentially localized Wannier orbitals for the top two moir\'e valence bands. Nevertheless, we examine the momentum-space weight of the localized orbital and the quantum geometry of the continuum band to justify the effective single-orbital reduction.

To quantify the validity of the single-orbital approximation, we compute the momentum-resolved localized-orbital weight $Z_f(\mathbf{k})$, which measures the fraction of the low-energy spectral weight carried by the partially Wannierized localized orbital. Using the notations defined in Eq.~(\ref{eq:udis_f}), it is given by the overlap between the active continuum-model Bloch state (the uppermost valence band, $n=1$) and the Bloch superposition of the localized Wannier orbital:
\begin{equation}
    Z_f(\mathbf{k}) = \left| \braket{u_{1\mathbf{k}} | \widetilde{u}_{f\mathbf{k}}} \right|^2.
\end{equation}
As shown in Fig.~\ref{fig:topo_info}(a), $Z_f(\mathbf{k})$ remains close to unity along the entirety of the half-filled Fermi surface. This indicates that the low-energy quasiparticles participating directly in the superconducting instability are strongly dominated by the localized orbital. The nonlocal component associated with the topological Wannier obstruction is concentrated away from the Fermi surface, primarily in the vicinity of the moir\'e $\gamma$ point. Since the Mott physics and the superexchange-driven pairing considered here are controlled predominantly by states near the Fermi surface, this provides direct microscopic support for the effective one-orbital description.

\begin{figure}[!ht]
    \centering
    \includegraphics[width=0.9\linewidth]{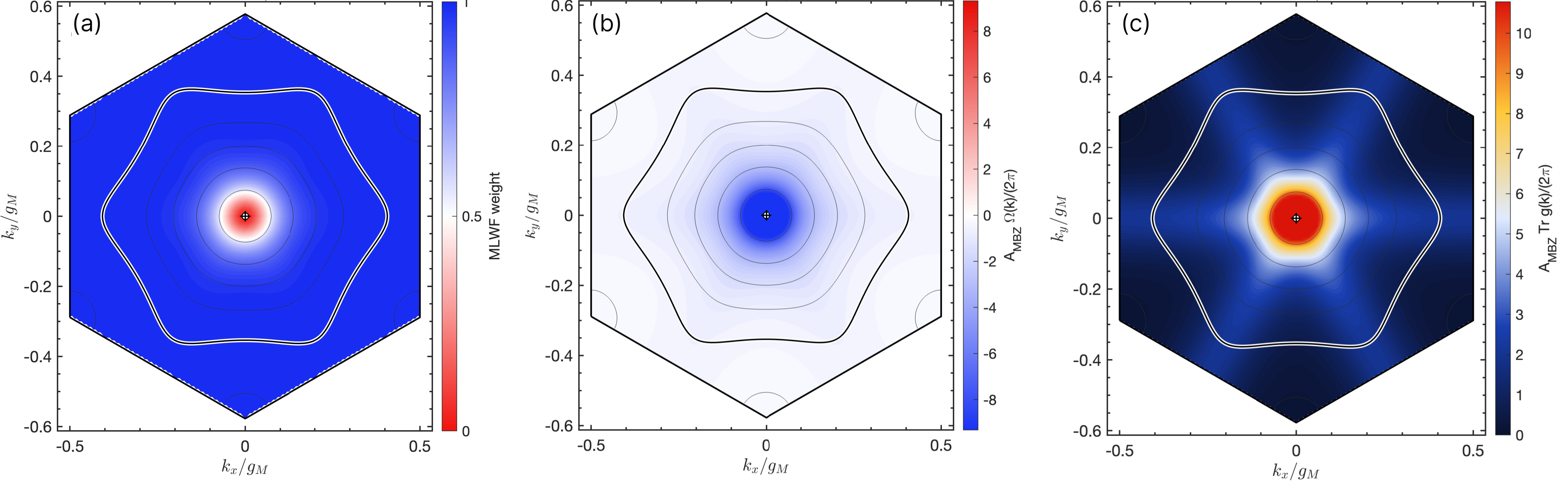}
    \caption{
    Momentum-space check of the localized-orbital description and quantum geometry. The moir\'e Fermi surface at half-filling is denoted by solid lines. 
    (a) The momentum-resolved localized-orbital weight $Z_f(\mathbf{k})$ overlaid in the moir\'e Brillouin zone. The color intensity indicates that the active states near the Fermi surface are predominantly of localized-orbital character. 
    (b) The normalized Berry curvature $\Omega(\mathbf{k})$ of the active continuum band. 
    (c) The normalized trace of the quantum metric $\mathrm{Tr}[g_{ij}(\mathbf{k})]$. Both quantum geometric quantities are sharply peaked near the $\gamma$ point and strongly suppressed along the Fermi surface.
    }
    \label{fig:topo_info}
\end{figure}

To further assess whether quantum geometric effects and multiband contributions might significantly alter the low-energy pairing physics~\cite{Peotta2015}, we evaluate the momentum-space distribution of the Berry curvature and the quantum metric for the active band. For the cell-periodic Bloch states $\ket{u_{1\mathbf{k}}}$, the quantum geometric tensor (QGT) is defined as
\begin{equation}
    Q_{\mu\nu}(\mathbf{k}) = \bra{\partial_{k_\mu} u_{1\mathbf{k}}} \left( 1 - \ket{u_{1\mathbf{k}}}\bra{u_{1\mathbf{k}}} \right) \ket{\partial_{k_\nu} u_{1\mathbf{k}}}.
\end{equation}
The Berry curvature and the quantum metric are given by the imaginary and real parts of the QGT, respectively:
\begin{align}
    \Omega(\mathbf{k}) &= -2 \, \mathrm{Im} \left[ Q_{xy}(\mathbf{k}) \right], \\
    g_{\mu\nu}(\mathbf{k}) &= \mathrm{Re} \left[ Q_{\mu\nu}(\mathbf{k}) \right].
\end{align}
The calculated Berry curvature $\Omega(\mathbf{k})$ and the trace of the quantum metric $\mathrm{Tr}[g_{ij}(\mathbf{k})]$ are presented in Figs.~\ref{fig:topo_info}(b) and \ref{fig:topo_info}(c). Both quantities are sharply peaked near the $\gamma$ point but remain relatively small along the active Fermi surface. 

In ideal flat-band systems where the Fermi velocity vanishes, the superfluid weight is completely dominated by the geometric contribution proportional to the quantum metric. However, in the intermediate-coupling regime considered here, the active band has an appreciable finite bandwidth and a sizable Fermi velocity along the half-filled Fermi surface. Because the low-energy Bogoliubov quasiparticles carry little Berry curvature and small quantum metric, and the $d+id$ pairing form factor has nodes near the $\gamma$ point where the metric is finite, the conventional intraband Fermi-velocity contribution is expected to dominate the superfluid response. Multiband mixing and quantum-geometric effects thus serve as subleading corrections to the pairing physics.

To further quantify the role of quantum geometry in the superconducting response, we estimate the relative importance of the conventional and geometric contributions to the superfluid weight. The total superfluid weight can be conceptually separated into a conventional intraband contribution and a geometric interband contribution, $D_{s,ab} = D^{\mathrm{conv}}_{s,ab} + D^{\mathrm{geom}}_{s,ab}$. For a dispersive band with a momentum-dependent pairing gap $\Delta_{\mathbf{k}}$, these two components scale as:
\begin{align}
    D^{\mathrm{conv}}_{s,ab} &\propto \sum_{\mathbf{k}} \frac{|\Delta_{\mathbf{k}}|^2}{E_{\mathbf{k}}^3} v_{a}(\mathbf{k}) v_{b}(\mathbf{k}), \\
    D^{\mathrm{geom}}_{s,ab} &\propto \sum_{\mathbf{k}} \frac{|\Delta_{\mathbf{k}}|^2}{E_{\mathbf{k}}} g_{ab}(\mathbf{k}),
\end{align}
where $v_{a}(\mathbf{k})$ is the band velocity along direction $a$, $E_{\mathbf{k}} = \sqrt{\xi_{\mathbf{k}}^2 + |\Delta_{\mathbf{k}}|^2}$ is the Bogoliubov quasiparticle energy, and $g_{ab}(\mathbf{k})$ is the quantum metric of the active band. The overall amplitudes of both terms are subject to the same Gutzwiller renormalization factors in the strong-coupling regime, but their relative importance is dictated by the momentum-space integrals.

For the chiral $d+id$-wave state on the triangular lattice, the pairing form factor is given by
\begin{equation}
    \Delta_{\mathbf{k}} \sim \Delta_d \left[ \cos(\mathbf{k} \cdot \bm{b}_1) + e^{i2\pi/3} \cos(\mathbf{k} \cdot \bm{b}_2) + e^{-i2\pi/3} \cos(\mathbf{k} \cdot \bm{b}_3) \right].
\end{equation}
Here $\bm{b}_m$ denotes the lattice vector of the moire triangular lattice; see main text.
The conventional term $D^{\mathrm{conv}}_{s,ab}$ is predominantly supported near the Fermi surface, where $E_{\mathbf{k}}$ reaches its minimum (proportional to the gap) and the band velocity remains sizable. By contrast, the geometric term $D^{\mathrm{geom}}_{s,ab}$ requires both a large quantum metric $g_{ab}(\mathbf{k})$ and a non-vanishing gap amplitude $|\Delta_{\mathbf{k}}|^2$. As shown in Fig.~\ref{fig:topo_info}(c), the quantum metric is strongly suppressed along the active Fermi surface. While $g_{ab}(\mathbf{k})$ is sharply peaked near the moir\'e $\gamma$ point ($\mathbf{k}=0$), the $d+id$-wave form factor strictly vanishes at this exact location, i.e., $\Delta_{\mathbf{k}=0} \propto 1 + e^{i2\pi/3} + e^{-i2\pi/3} = 0$. Consequently, the geometric integrand $|\Delta_{\mathbf{k}}|^2 g_{ab}(\mathbf{k}) / E_{\mathbf{k}}$ is strongly suppressed everywhere in the Brillouin zone. The superconducting response is thus definitively dominated by the conventional Fermi-velocity contribution, further confirming that quantum geometric effects are a subleading correction to the gossamer superconductivity in this system.

\subsection{A4. Effect of displacement fields}

While the twist angle serves as a primary tuning parameter, an experimentally important and highly accessible tuning knob is the out-of-plane displacement field. Experiments have shown that a displacement field applied at half-filling ($\nu = 1$) can systematically suppress superconductivity in favor of a correlated insulating state, with signs of antiferromagnetic (AFM) order at large fields. In this section, we provide a phenomenological theoretical understanding of this displacement-field-driven transition by bridging the continuum model to our effective lattice description.

In the continuum model, the displacement field enters as a layer potential difference. The Hamiltonian for the top moir\'e valence bands near the $K_+$ valley is given by
\begin{equation}
    H_{0,\uparrow}(\mathbf{k},\mathbf{r}; V_z) =
    \begin{bmatrix}
    h_b(\mathbf{k},\mathbf{r}) + V_z/2 & \Delta_T(\mathbf{r}) \\
    \Delta_T^\dagger(\mathbf{r}) & h_t(\mathbf{k},\mathbf{r}) - V_z/2
    \end{bmatrix}.
\end{equation}
As shown in Eq.~\eqref{eq:continuum_model}, $h_b$ and $h_t$ represent the intralayer kinetic and moir\'e potential terms for the bottom and top layers respectively, $V_z$ is the layer potential difference proportional to the screened displacement field, and the Hamiltonian in the opposite valley is obtained by time-reversal symmetry.

After Wannierization, this layer asymmetry produces an effective triangular-lattice model with spin-dependent complex hopping phases:
\begin{equation}
    \mathcal{H}_0(V_z) = -\sum_{n} \sum_{\langle ij \rangle_n, \sigma} t_n(V_z) e^{i\sigma \phi_{ij}^{(n)}(V_z)} c_{i\sigma}^\dagger c_{j\sigma} + \text{h.c.},
\end{equation}
where $\sigma=\pm 1$ denotes the spin-valley index, $t_n(V_z)$ are the hopping amplitudes depending on the displacement field $V_z$, and $\phi_{ij}^{(n)}$ are the bond-dependent complex phases. The sign convention for these phases is defined according to the representative positive bond directions, e.g., $\phi_{ij}^{(n)} = \phi_n$ if the bond $i \rightarrow j$ is along the direction of the colored arrows, and $\phi_{ij}^{(n)} = -\phi_n$ otherwise, as schematically illustrated in Fig.~\ref{fig:dfield_phase_schematic}. 

\begin{figure}[!ht]
    \centering
    \includegraphics[width=0.22\linewidth]{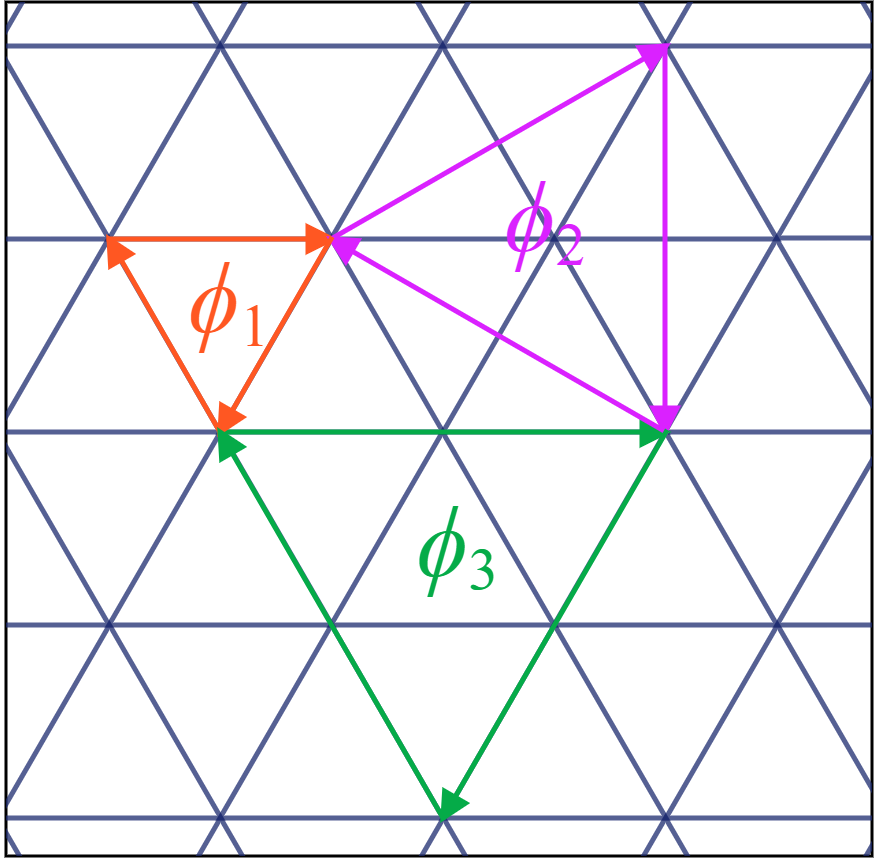}
    \caption{
    The phases $\phi_1$, $\phi_2$, and $\phi_3$ denote the displacement-field-induced phases associated with the 1st-, 2nd-, and 3rd-neighbor hoppings, respectively. The colored arrows indicate the representative positive bond directions used to define the sign convention of $\phi_{ij}^{(n)}$. For instance, $\phi_{ij}^{(n)}=\phi_n$ when bond $ij$ is along the direction of the colored arrow, and $\phi_{ij}^{(n)}=-\phi_n$ otherwise.
    }
    \label{fig:dfield_phase_schematic}
\end{figure}

The on-site Hubbard term $U$ is almost unchanged with varying $V_z$. In the strong-coupling language, the nearest-neighbor complex hopping generates an anisotropic exchange interaction with a Dzyaloshinskii--Moriya (DM) component:
\begin{equation}
    \mathcal{H}_J(V_z) = J \sum_{\langle ij \rangle} \left[ S_i^z S_j^z + \cos(2\phi_{ij}) \left(S_i^x S_j^x + S_i^y S_j^y\right) + \sin(2\phi_{ij}) \hat{z} \cdot (\mathbf{S}_i \times \mathbf{S}_j) \right].
\end{equation}
Thus, a displacement field moves the system out of the two-parameter real-hopping phase diagram and introduces an additional physically important axis, namely the complex hopping/DM phase.

We further examine how the single-orbital tight-binding model evolves under a displacement field $V_z$, using both the trivial- and topological-band continuum-model parameters. The results are summarized in Figs.~\ref{fig:dfield_trivial} and \ref{fig:dfield_topo}. In both cases, we find that the first-nearest-neighbor hopping magnitude $|t_1|$ increases with increasing $V_z$. Since the effective Hubbard interaction $U$ is mainly controlled by the spatial extent of the localized moir\'e orbital and changes only weakly with $V_z$, the relative interaction strength $U/|t_1|$ decreases as $V_z$ increases. Therefore, the experimentally observed suppression of SC by a displacement field is not naturally interpreted in our model as a simple displacement-field-driven increase of $U/|t_1|$ into the large-$U$ Mott regime.

\begin{figure}[!ht]
    \centering
    \includegraphics[width=0.8\linewidth]{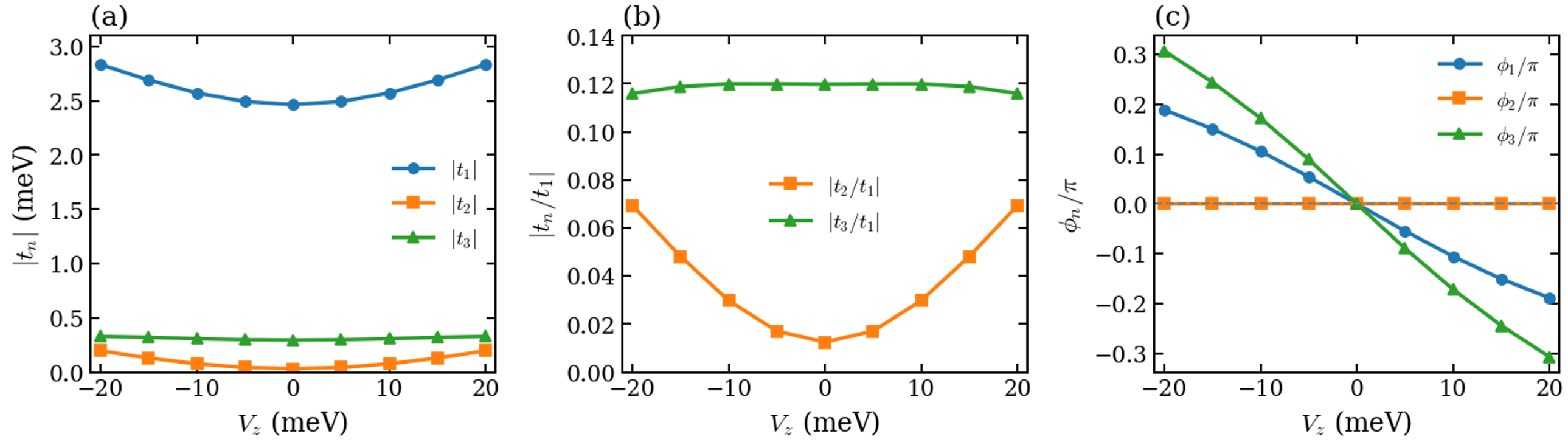}
    \caption{
    Displacement-field effect for the trivial-band continuum-model parameters. The conversion between the screened layer bias $E$ and $V_z$ is approximately $100 \text{ mV nm}^{-1} \simeq 26 \text{ meV}$. (a) Magnitudes of the hopping amplitudes $|t_1|$, $|t_2|$, and $|t_3|$ as functions of $V_z$. (b) Hopping ratios $|t_2|/|t_1|$ and $|t_3|/|t_1|$. (c) Hopping phases $\phi_n/\pi$ as functions of $V_z$.
    }
    \label{fig:dfield_trivial}
\end{figure}

\begin{figure}[!ht]
    \centering
    \includegraphics[width=0.8\linewidth]{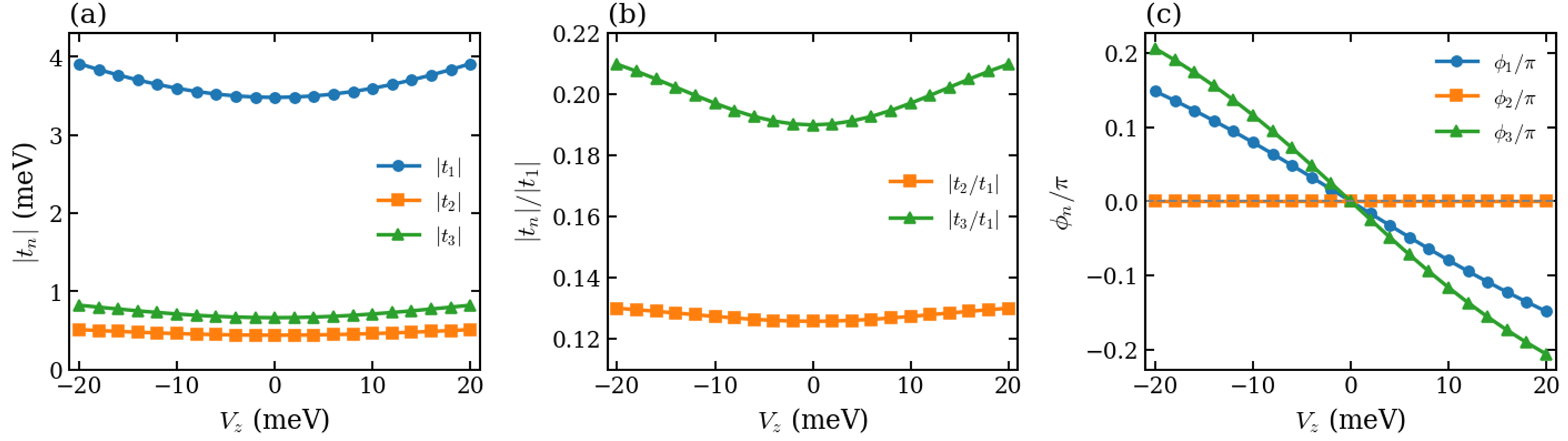}
    \caption{
    Similar to Fig.~\ref{fig:dfield_trivial}, displacement-field effect for the topological-band continuum-model parameters.
    }
    \label{fig:dfield_topo}
\end{figure}

Instead, the more important effect of $V_z$ is the rapid development of spin-dependent complex hopping phases $\phi_n$, as shown in Figs.~\ref{fig:dfield_trivial}(c) and \ref{fig:dfield_topo}(c). These phases modify the band dispersion, shift the van-Hove structure and Fermi-surface geometry, and generate anisotropic exchange and DM interactions in the strong-coupling limit. Such a DM interaction associated with complex phases breaks the SU(2) spin rotational symmetry and is in favor of a classical in-plane $120^\circ$ N\'eel order. Moreover, it may activate triplet SC channels such as $p$-wave pairings, which we leave for further study.

To examine the effect of the displacement-field-induced complex hopping within a common variational framework, we fix $U/t_1 = 8$, $t_2/t_1 = 0.1$, $t_3/t_1 = 0.2$, and $J/t_1 = 0.3$ at half filling, and compare the fully optimized energies of three competing Gutzwiller-projected states: the gossamer SC state, the correlated normal state, and the $120^\circ$ AFM-ordered state. For simplicity, we take $\phi_2 = \sqrt{3}\phi_1$ and $\phi_3 = 2\phi_1$, consistent with phases proportional to the corresponding bond lengths.

\begin{figure}[!ht]
    \centering
    \includegraphics[width=0.6\linewidth]{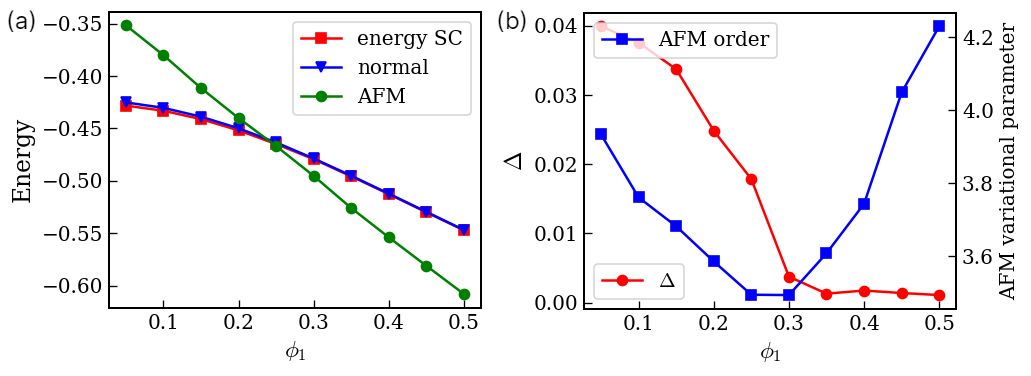}
    \caption{
    (a) The per-site energy of projected gossamer SC states, projected normal states, and projected AFM ordered state as functions of the first-neighbor hopping phase $\phi_1$. (b) The variational SC gap $\Delta$ and variational AFM parameter $m$ as functions of $\phi_1$. Here we use $U/t_1 = 8$, $t_2/t_1 = 0.1$, $t_3/t_1 = 0.2$, and $J/t_1 = 0.3$ at half filling. For simplicity, we use $\phi_2 = \sqrt{3}\phi_1$ and $\phi_3 = 2\phi_1$.
    }
    \label{fig:dfield_energy_afm}
\end{figure}

As shown in Fig.~\ref{fig:dfield_energy_afm}(a), the gossamer SC state has the lowest variational energy for small $\phi_1$. As $\phi_1$ increases, the energy of the AFM-ordered state decreases more rapidly and crosses that of the SC state near $\phi_1 \simeq 0.25$, signaling a transition from the gossamer SC state to an AFM-ordered ground state. At the same time, the optimized pairing amplitude $\Delta$ is strongly suppressed and eventually approaches zero, as shown in Fig.~\ref{fig:dfield_energy_afm}(b). Consequently, the optimized SC and projected-normal-state energies become indistinguishable at larger $\phi_1$. These results demonstrate that the complex hopping phases induced by a displacement field destabilize gossamer superconductivity and favor antiferromagnetic order, providing a natural explanation for the experimentally observed replacement of the superconducting state by a correlated insulating state. Note that the variational AFM parameter is large which can fully gap the original Fermi surface.

This theoretical picture is directly supported by experiment: in tWSe$_2$, applying a displacement field at $\nu = 1$ systematically suppresses the superconducting state and stabilizes a correlated insulating phase. This trend is consistent with previous Hartree-Fock studies of the moir\'e Hubbard model~\cite{Zang2021_HF}, which showed that complex hopping phases and the associated changes in the van-Hove structure can strongly enhance magnetic insulating instabilities with a full charge gap. Therefore, a displacement field can suppress the gossamer superconducting state in favor of a correlated magnetic insulating state, such as a Slater-type insulator with large in-plane N\'eel order.

\section{B. Details of the Renormalized Mean-Field Theory}\label{app:RMFT}

As outlined in the main text, we evaluate the ground-state energy of the $t_1$-$t_2$-$t_3$ extended Hubbard model using the Gutzwiller-projected wavefunction $|\Psi_G\rangle = P_G |\Psi_{\text{MF}}\rangle$. In this Appendix, we provide the rigorous analytical details of the Gutzwiller approximation, which maps the intractable projected energy functional $E_G$ onto a computationally tractable problem defined on the unprojected mean-field state $|\Psi_{\text{MF}}\rangle$.

We first evaluate the normalization denominator $\mathcal{N} = \langle \Psi_{\text{MF}} | P_G^2 | \Psi_{\text{MF}} \rangle$ in the physical double occupancy $d_i = \langle \Psi_G | n_{i\uparrow}n_{i\downarrow} | \Psi_G \rangle / \langle \Psi_G | \Psi_G \rangle$.  Since $P_{G,i}^2 = 1 - (1-g_i^2) n_{i\uparrow}n_{i\downarrow}$, evaluating $\mathcal{N}$ exactly requires dealing with spatial correlations of all orders. Here, we introduce the tree-level approximation, which neglects non-local spatial correlations encoded in loops. The expectation value of the product is then simply factorized into the product of local expectation values:
\begin{equation}
    \mathcal{N} \approx \prod_i \left( 1 - (1-g_i^2) d_{0,i} \right),
\end{equation}
where $d_{0,i} = \langle n_{i\uparrow}n_{i\downarrow} \rangle_0$ is the bare double occupancy evaluated with respect to the unprojected state $|\Psi_{\text{MF}}\rangle$. 
Applying the same tree-level approximation to the numerator $\langle \Psi_{\text{MF}} | P_G n_{i\uparrow}n_{i\downarrow} P_G | \Psi_{\text{MF}} \rangle$, we isolate the operator at site $i$ from the rest of the lattice. Noting that $P_{G,i}^2 n_{i\uparrow}n_{i\downarrow} = g_i^2 n_{i\uparrow}n_{i\downarrow}$, we obtain
\begin{equation}
    d_{i} = \frac{g_i^2 d_{0,i}}{1 - (1-g_i^2)d_{0,i}}.
    \label{eq:app_di}
\end{equation}
As emphasized in the main text, projecting out $d_i$ inevitably redistributes the probabilities of the other local configurations. To preserve the local total particle number $n_{i} = \langle n_{i} \rangle_G \approx \langle n_{i} \rangle_0 \equiv n_{0,i}$, the statistical probabilities of the four possible local configurations (empty, singly occupied by spin-up, singly occupied by spin-down, and doubly occupied) at site $i$ are strictly constrained. Denoting the unprojected local density as $n_{0,i\sigma} = \langle n_{i\sigma} \rangle_0$, the renormalized probabilities are given by:
\begin{eqnarray}
    e_{i} &=& \langle (1-n_{i\uparrow})(1-n_{i\downarrow}) \rangle_G \approx 1 - n_{0,i} + d_i, \\
    m_{i\sigma} &=& \langle n_{i\sigma}(1-n_{i\bar{\sigma}}) \rangle_G \approx n_{0,i\sigma} - d_i, \\
    d_{i} &=& \langle n_{i\uparrow}n_{i\downarrow} \rangle_G \approx \frac{g_i^2 d_{0,i}}{1 - (1-g_i^2)d_{0,i}}.
\end{eqnarray}
These four probabilities exactly sum to unity ($e_i + m_{i\uparrow} + m_{i\downarrow} + d_i = 1$) and form the foundation for renormalizing the non-local operators.

The suppression of double occupancies fundamentally alters the effective amplitude of inter-site physical processes. We capture this by mapping the physical operators onto the unprojected state, dressed with renormalization factors. 

\textbf{Kinetic energy ($\tilde{\mathcal{H}}_0$).} 
The inter-site electron hopping $c_{i\sigma}^\dagger c_{j\sigma}$ alters the local configurations at both site $i$ (arrival) and site $j$ (departure). Based on the altered phase space probabilities, the physical hopping expectation value is approximated as $\langle c_{i\sigma}^\dagger c_{j\sigma} \rangle_G \approx q_{ij\sigma} \langle c_{i\sigma}^\dagger c_{j\sigma} \rangle_0$. The renormalization factor $q_{ij\sigma}$ is the ratio of the projected transition probability to the bare one:
\begin{equation}
\label{eq:q_t}
q_{ij\sigma} = \frac{
    \left( \sqrt{m_{i\sigma} e_i} + \sqrt{d_i m_{i\bar{\sigma}}} \right)
    \left( \sqrt{m_{j\sigma} e_j} + \sqrt{d_j m_{j\bar{\sigma}}} \right)
}{
    \left( \sqrt{m_{0,i\sigma} e_{0,i}} + \sqrt{d_{0,i} m_{0,i\bar{\sigma}}} \right)
    \left( \sqrt{m_{0,j\sigma} e_{0,j}} + \sqrt{d_{0,j} m_{0,j\bar{\sigma}}} \right)
}.
\end{equation}
Consequently, the effective kinetic Hamiltonian takes the form:
\begin{equation}
    \tilde{\mathcal{H}}_0 = -\sum_{n=1,2,3}\sum_{\langle i j \rangle_n, \sigma} t_n q_{ij\sigma} \left(c^\dagger_{i, \sigma} c_{j, \sigma} + \text{h.c.}\right).
\end{equation}

\textbf{Exchange energy ($\tilde{\mathcal{H}}_J$).}
The superexchange interaction $H_J = J \sum_{\langle ij \rangle_1} \mathbf{S}_i \cdot \mathbf{S}_j$ acts exclusively on singly occupied sites. Because the projection partially removes doublons and holes, it effectively enhances the probability of finding a localized spin. The spin operator is thus renormalized by the ratio of singly-occupied probabilities, $\mathbf{S}_i \rightarrow g_{s,i} \mathbf{S}_i$, where $g_{s,i} = (m_{i\uparrow} + m_{i\downarrow}) / (m_{0,i\uparrow} + m_{0,i\downarrow})$. 
The effective exchange Hamiltonian becomes:
\begin{equation}
    \tilde{\mathcal{H}}_J = J \sum_{\langle i j \rangle_1} g_{s,i} g_{s,j} \mathbf{S}_i \cdot \mathbf{S}_j.
\end{equation}
Note that for terms breaking SU(2) symmetry (e.g., transverse vs. longitudinal spin components), the renormalization factors can be generalized respectively depending on the local magnetization.

Combining the above, the total variational ground-state energy reduces to a simple function of $g_i$ and the unprojected mean-field densities:
\begin{equation}
    E_G(g_i, \{\alpha\}) \approx \langle \Psi_{\text{MF}}(\{\alpha\}) | \tilde{\mathcal H}_0 + \tilde{\mathcal H}_J | \Psi_{\text{MF}} (\{\alpha\})\rangle + U \sum_i d_i.
\end{equation}
The optimal ground state is found by minimizing $E_G$ with respect to both $g_i$ and the mean-field parameters $\{\alpha\}$ in $| \Psi_{\text{MF}}\rangle$.

\textbf{Renormalization of the SC pairing.}
We distinguish the dimensionless anomalous pair expectation value from the pairing parameter $\Delta_{ij}$ entering the mean-field Hamiltonian, which has units of energy. Defining the projected and unprojected pair expectation values as $\Phi^{\mathrm{phys}}_{ij}=\langle c_{i\uparrow}^\dagger c_{j\downarrow}^\dagger\rangle_G$ and $\Phi^{\mathrm{MF}}_{ij}=\langle c_{i\uparrow}^\dagger c_{j\downarrow}^\dagger\rangle_0$, respectively, the Gutzwiller approximation gives
\begin{equation}
    \Phi^{\mathrm{phys}}_{ij} \approx q_{ij}\Phi^{\mathrm{MF}}_{ij},
\end{equation}
where $q_{ij\uparrow}=q_{ij\downarrow}\equiv q_{ij}$ has been assumed for a paramagnetic background. For a translationally invariant system at half-filling, the suppression of double occupancy gives $q_{ij}\propto g^2$~\cite{Zhang2003}. In the main text, we apply this leading Gutzwiller renormalization to the energy-valued mean-field pairing parameter and denote the resulting physical pairing scale by $g^2\Delta$; it should not be identified directly with the dimensionless expectation value $\Phi^{\mathrm{phys}}_{ij}$.

\section{C. Mean-Field Ans\"atze}
\label{app:ansatzes}

By specifying the variational parameters $(\chi_{ij}, \Delta_{ij}, \mathbf{m}_{i})$ used in the mean-field Hamiltonian in the main text, we construct the following trial states. We parameterize the lattice sites as $\mathbf{r} = n\mathbf{b}_1 + m\mathbf{b}_2$ and denote the three primitive nearest-neighbor (NN) bond directions as $\mathbf{b}_1$, $\mathbf{b}_2$, and $\mathbf{b}_3 = \mathbf{b}_1 - \mathbf{b}_2$.

{\bf Normal state.}
The uniform normal state preserves all lattice symmetries without any symmetry-breaking orders:
\begin{equation}
    \Delta_{ij} = 0, \quad \mathbf{m}_i = \mathbf{0}, \quad \chi_{\mathbf{r}, \mathbf{r}+\mathbf{b}_m} = \chi_1.
\end{equation}
Further-neighbor hoppings ($\chi_2$ and $\chi_3$) are similarly kept as uniform, real amplitudes.

When a finite displacement field is applied, the spin-valley-dependent complex phases generated in the microscopic hopping amplitudes are incorporated consistently into the variational hopping ansatz. Specifically, for an $n$th-neighbor bond $ij$, we generalize the hopping parameter as $\chi_{ij\sigma}=\chi_n e^{i\sigma\phi_{ij}^{(n)}}$, using the same bond-dependent phase $\phi_{ij}^{(n)}$ as in the effective kinetic Hamiltonian of Sec.~A4. Therefore, only the amplitudes $\chi_n$ remain variational parameters, while the complex phases are fixed.

{\bf Superconducting states.}
The SC states emerge from the NS ($\mathbf{m}_i = \mathbf{0}$) by introducing a finite NN spin-singlet pairing $\Delta_{\mathbf{r}, \mathbf{r}+\mathbf{b}_m}$. Based on the irreducible representations of the triangular lattice, we classify them as:
\begin{itemize}
    \item \textbf{Extended $s$-wave:} Isotropic pairing on all bonds,
    \begin{equation}
        \Delta_{\mathbf{r}, \mathbf{r}+\mathbf{b}_m} = \Delta_s, \quad (m=1,2,3).
    \end{equation}
    \item \textbf{Nematic $d$-wave:} Pairing breaks the $C_6$ rotational symmetry, spanned by two degenerate components:
    \begin{align}
        \Delta^{(d_1)}_{\mathbf{r}, \mathbf{r}+\mathbf{b}_m} &= \Delta_d \cos\left[\frac{2\pi(m-1)}{3}\right], \\
        \Delta^{(d_2)}_{\mathbf{r}, \mathbf{r}+\mathbf{b}_m} &= \Delta_d \sin\left[\frac{2\pi(m-1)}{3}\right].
    \end{align}
    \item \textbf{Chiral $d+id$-wave:} A time-reversal symmetry-breaking linear combination:
    \begin{equation}
        \Delta^{(d+id)}_{\mathbf{r}, \mathbf{r}+\mathbf{b}_m} = \Delta_d \exp\left[i\frac{2\pi(m-1)}{3}\right].
    \end{equation}
\end{itemize}

{\bf Magnetic state with staggered flux.}
To capture the strongly correlated regime, we introduce a mean-field ansatz that doubles the unit cell to accommodate an emergent gauge flux $\varphi$ and a $120^\circ$ coplanar antiferromagnetic order.
\begin{itemize}
    \item \textbf{Hopping ansatz:} The gauge flux modulates the NN hopping spatially along the $\mathbf{b}_1$ direction (characterized by the coordinate $n$), explicitly doubling the unit cell:
\begin{align}
    \chi_{\mathbf{r},\mathbf{r}+\mathbf{b}_1,\sigma} &= \chi_1, \\
    \chi_{\mathbf{r},\mathbf{r}+\mathbf{b}_2,\sigma} &= \chi_1 (-1)^n, \\
    \chi_{\mathbf{r},\mathbf{r}+\mathbf{b}_3,\sigma} &= \chi_1 (-1)^n.
\end{align}
This alternating sign yields a zero flux in the up-triangles and $\pi$ flux in the down-triangles.
    \item \textbf{Magnetic order:} The local Weiss field $\mathbf{m}_{\mathbf{r}}$ characterizes the $120^\circ$ N\'eel order,
    \begin{equation}
        \mathbf{m}_{\mathbf{r}} = m \left( \cos(\mathbf{Q} \cdot \mathbf{r}),\quad \sin(\mathbf{Q} \cdot \mathbf{r}),\quad 0 \right),
    \end{equation}
    where $m$ is the magnetization amplitude and $\mathbf{Q} = (4\pi/3, 0)$ is the ordering wavevector defining the three sublattices.
\end{itemize}

\section{D. Additional Details of the RMFT Results}
\label{app:rmft_details}

This section provides additional RMFT results concerning the roles of the superexchange interaction, the extended hoppings, doping, and dielectric screening in determining the stability of the gossamer superconducting state.

\subsection{D1. Role and Parameter Dependence of the Superexchange $J$}
\label{app:J_term}

Within the present RMFT formulation, $U$ and $J$ play distinct roles. The Hubbard interaction controls the energetic cost of double occupancy and thereby the optimized Gutzwiller factor $g$, whereas the antiferromagnetic exchange $J$ describes the virtual hopping process. The gossamer SC state found at intermediate coupling therefore combines residual charge fluctuations ($0<g<1$) with exchange-driven pairing. Within this approximation, an explicit finite $J$ is required to obtain this superconducting solution. This methodological requirement does not imply that the microscopic system must contain an exchange interaction independent of $U$: in a pure Hubbard model, antiferromagnetic exchange can arise dynamically from virtual hopping processes, which are not generated automatically by the present RMFT treatment.

We keep $J/t_1=0.25$ fixed in the main calculations to isolate the evolution driven by the charge-correlation scale $U/t_1$, rather than varying the charge and exchange scales simultaneously. This is a modeling choice, not a microscopic relation assumed over the full interaction range. The explicit $\mathcal H_J$ should be understood as part of the present RMFT energy functional rather than as an additional microscopic interaction independent of $U$. In this implementation, $\mathcal H_U$ enters through the projected double occupancy, and the approximation does not separately generate a virtual-exchange term $4t_1^2/U$ from $\mathcal H_U$. We therefore treat $\mathcal H_J$ as the exchange contribution absent from this approximation, rather than as a second evaluation of an exchange contribution already included through $\mathcal H_U$.

\begin{figure}[!ht]
    \centering
    \includegraphics[width=0.7\linewidth]{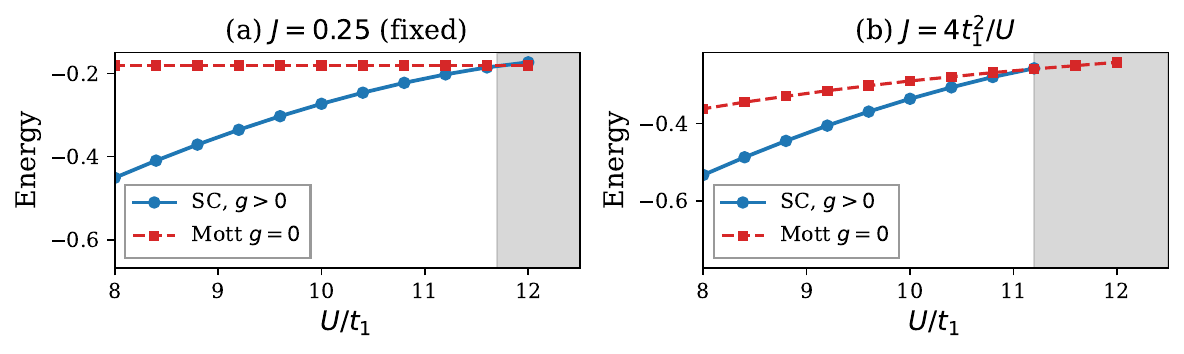}
    \caption{
    Comparison of the variational energies of the gossamer SC state and the Mott insulating state as functions of $U/t_1$. 
    (a) Using a fixed superexchange $J/t_1 = 0.25$, as employed in the main text. 
    (b) Using the strong-coupling scaling relation $J = 4t_1^2/U$. 
    Both calculations are performed with $t_2/t_1 = 0$ and $t_3/t_1 = 0.2$ at half-filling. The qualitative structure of the phase diagram remains unchanged.
    }
    \label{fig:j_scaling}
\end{figure}

To assess the sensitivity to the fixed-$J$ choice, we additionally impose the strong-coupling estimate $J=4t_1^2/U$ for $U/t_1\geq8$. As shown in Fig.~\ref{fig:j_scaling}, the intermediate gossamer-SC regime remains stable, while the SC--Mott boundary shifts only modestly, from $U_c/t_1\approx11.6$ for fixed $J$ to $U_c/t_1\approx11.2$ for $J=4t_1^2/U$. The qualitative phase evolution is therefore robust against this alternative parameterization of $J$. We note that the effective one-orbital model is obtained from a multiorbital continuum description, so its exchange interaction need not be exhausted by the simple single-band estimate $4t_1^2/U$. Whether the same gossamer phase survives in a controlled treatment of the pure Hubbard model at $J=0$ remains an open question requiring methods that retain the virtual exchange processes explicitly.

\subsection{D2. Effects of the Extended Hoppings $t_2$ and $t_3$}
\label{app:extended_hoppings}

The extended hoppings $t_2$ and $t_3$ do not introduce additional attractive interactions: the spin-singlet pairing channel in our calculation continues to originate from the nearest-neighbor antiferromagnetic exchange $J$. Their role is instead to modify the kinetic-energy structure of the partially projected state. To complement the colormaps in Fig.~3 of the main text, Figs.~\ref{fig:g_linecuts}(b) and \ref{fig:g_linecuts}(c) show representative line cuts of the optimized Gutzwiller factor and projected pairing amplitude as functions of $t_3/t_1$ for selected $t_2/t_1$. The enhancement of the physical pairing amplitude over an intermediate range of $t_3/t_1$ does not arise from an increase in $g$, while sufficiently large $t_3/t_1$ eventually suppresses the pairing amplitude. The influence of the extended hoppings on superconductivity is therefore nonmonotonic.

\begin{figure}[!ht]
    \centering
    \includegraphics[width=\linewidth]{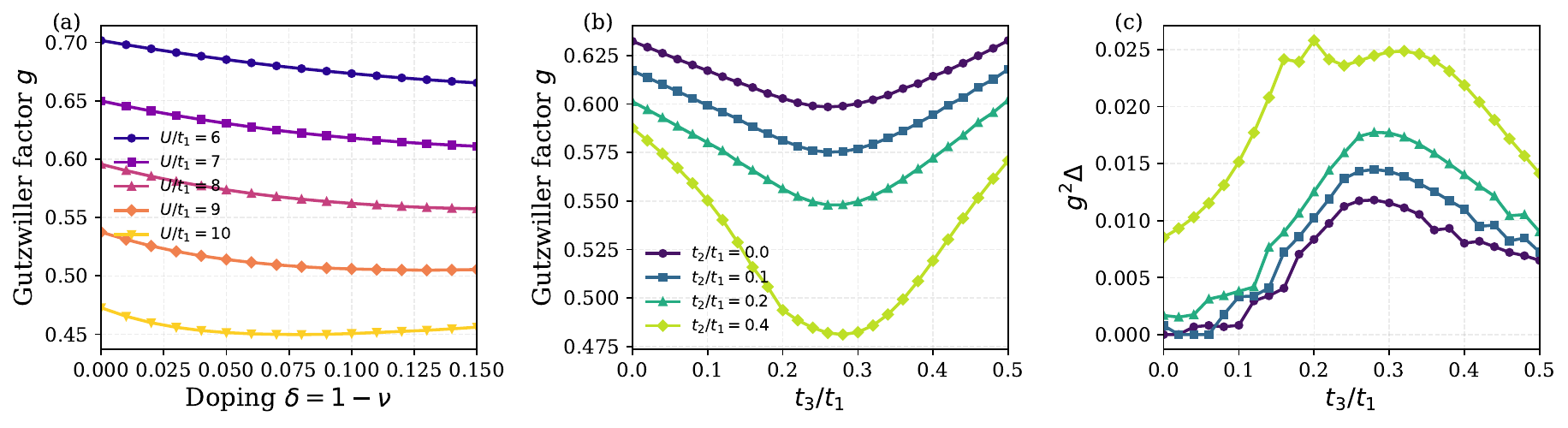}
    \caption{
    Line cuts of the optimized Gutzwiller factor and physical pairing amplitude corresponding to Fig.~3 of the main text.
    (a) Optimized Gutzwiller factor $g$ as a function of hole doping $\delta=1-\nu$ for selected values of $U/t_1$, with $t_2/t_1=0.05$, $t_3/t_1=0.18$, and $J/t_1=0.25$.
    (b) Optimized Gutzwiller factor $g$ and (c) physical pairing amplitude $g^2|\Delta|$ as functions of $t_3/t_1$ for selected values of $t_2/t_1$, with $U/t_1=8$ and $J/t_1=0.25$ at half filling.
    }
    \label{fig:g_linecuts}
\end{figure}

\subsection{D3. Doping Dependence and Suppression of Gossamer Superconductivity}
\label{app:doping_suppression}

In the main text, we noted that the gossamer SC pairing is peaked at half-filling ($\nu=1$) and is suppressed upon doping. Figure~\ref{fig:g_linecuts}(a) additionally shows the optimized Gutzwiller factor $g$ as a function of doping for several values of $U/t_1$. These line cuts demonstrate that the half-filled superconducting state remains partially projected, with a finite $g$, and provide a direct view of the evolution of charge renormalization away from half-filling. The rapid suppression of superconductivity is controlled by the energetic competition between the paired and correlated normal states, as clarified by the following energy-decomposition analysis.

The stability of the gossamer SC state is determined by an energetic competition. To see this, we define the total per-site energy difference between the optimized SC state and the optimized correlated normal state as $\Delta E_{\mathrm{tot}} = E_{\mathrm{SC}} - E_{\mathrm{Normal}}$. We can explicitly decompose this total energy difference into its kinetic, exchange, and Hubbard contributions:
\begin{equation}
    \Delta E_{\mathrm{tot}} = \Delta E_{\mathrm{kin}} + \Delta E_{J} + \Delta E_{U}.
\end{equation}
A negative $\Delta E_{\mathrm{tot}}$ indicates that the superconducting state is energetically favored.

At half filling, the SC state pays a kinetic-energy cost, $\Delta E_{\mathrm{kin}}>0$, but gains both exchange and Hubbard energies, $\Delta E_J<0$ and $\Delta E_U<0$. Their combined contribution overcomes the kinetic-energy cost and stabilizing the gossamer SC state. Upon hole doping, all three energy differences decrease in magnitude and approach zero; see Fig.~\ref{fig:energy_doping} (a). In particular, the exchange- and Hubbard-energy gains, $\Delta E_{J}$ and $\Delta E_{U}$, collapse rapidly. Although the kinetic-energy cost also decreases, the superconducting state progressively loses its correlation-energy advantage, and $\Delta E_{\mathrm{tot}}$ approaches zero as the optimized superconducting solution merges with the correlated normal state.

\begin{figure}[!ht]
    \centering
    \includegraphics[width=0.65\linewidth]{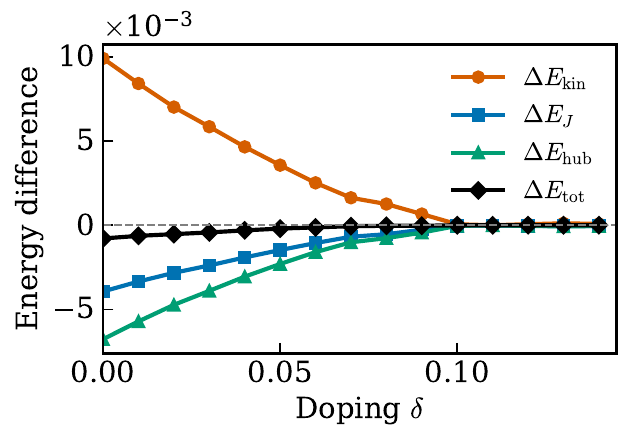}
    \caption{
    Energy decomposition of the gossamer SC state relative to the correlated normal state as a function of hole doping $\delta = 1 - \nu$, calculated at $U/t_1=8.4$, $t_2/t_1=0.05$, $t_3/t_1=0.18$, and $J/t_1=0.25$. The total variational energy difference $\Delta E_{\mathrm{tot}}$ is decomposed into kinetic ($\Delta E_{\mathrm{kin}}$), exchange ($\Delta E_J$), and Hubbard ($\Delta E_U$) contributions. The SC state is energetically stable only when $\Delta E_{\mathrm{tot}} < 0$, which occurs exclusively in a narrow window near half-filling. (b) The hopping renormalization factor $q_{t}$ and single-site spin renormalization factor $g_{s}$ as functions of hole doping; see definitions of $q_{t}$ and $g_{s}$ in Section~B. The model parameters are the same as those in (a). 
    }
    \label{fig:energy_doping}
\end{figure}

This evolution is accompanied by a decrease in the spin-renormalization factor $g_s$ and an enhancement of the hopping-renormalization factor $q_t$, as defined in Sec.~B. Figure~S7(b) shows $q_t$, which renormalizes the kinetic term according to
  \[
  t\left(c_{i\sigma}^{\dagger}c_{j\sigma}+\mathrm{h.c.}\right)
  \rightarrow
  q_t t\left(c_{i\sigma}^{\dagger}c_{j\sigma}+\mathrm{h.c.}\right).
  \]
The increase of $q_t$ reflects the redistribution of local configuration probabilities upon hole doping. 
In particular, the increased empty-site probability opens additional hopping channels, so that finite hole doping permits charge motion even in the limit $g=0$. By contrast, $g_s$ renormalizes the local spin operator as $\mathbf{S}_i\rightarrow g_s\mathbf{S}_i$. As shown in Fig.~S7(b), $g_s$ decreases with doping, and consequently the exchange-bond renormalization $g_s^2$ is reduced. Hole doping therefore enhances charge motion while weakening the exchange processes that support spin-singlet pairing, consistent with the rapid loss of the correlation-energy advantage of the superconducting state.

It is instructive to contrast this behavior with that of a conventional doped Mott insulator. In a rigid Mott state ($g=0$), charge carriers are completely frozen at half-filling, and finite doping is fundamentally required to mobilize them. The gossamer SC state, by contrast, already sustains a finite density of mobile doublons and holes at half-filling ($g>0$).  Doping therefore does not open a qualitatively new charge-coherence channel. Instead, it introduces a net density of holes while suppressing the exchange- and Hubbard-energy gains that stabilize the SC state. Consequently, the SC is intrinsically fragile against doping, with the pairing amplitude peaking sharply at $\nu=1$ and decaying rapidly away from half-filling. This behavior naturally explains the remarkably narrow SC phase observed experimentally in twisted WSe$_2$.

\subsection{D4. Robustness against Dielectric-Screening Uncertainty}
\label{app:epsilon_robustness}

The dielectric constant of twisted bilayer WSe$_2$ has not been directly measured. As a material reference, first-principles calculations for bilayer WSe$_2$ give $\epsilon_{\parallel}\simeq15.7$ and $\epsilon_{\perp}\simeq7.6$~\cite{Laturia2018}, corresponding to an effective value $\epsilon=\sqrt{\epsilon_{\parallel}\epsilon_{\perp}}\simeq10.9$. This motivates our representative choice $\epsilon\simeq10$, while retaining a quantitative uncertainty. Writing $U=U_0/\epsilon$, we test $\epsilon=8$--$12$ in Fig.~\ref{fig:epsilon_robustness}. Although screening shifts the estimated SC--Mott boundary, the intermediate gossamer-SC regime remains robust throughout this range.

Future thermodynamic-compressibility measurements could provide a more direct constraint on the effective interaction scale $U$ entering the one-orbital model and on the corresponding effective dielectric screening $\epsilon$.

\begin{figure}[!ht]
    \centering
    \includegraphics[width=0.55\linewidth]{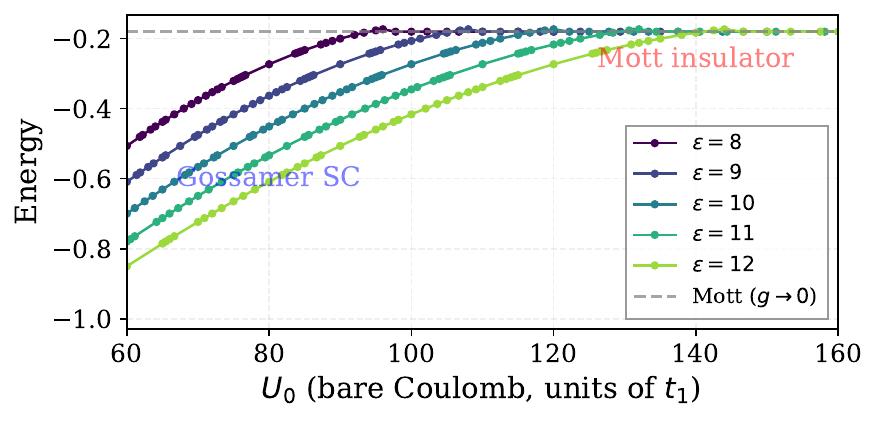}
    \caption{
    Sensitivity of the RMFT phase evolution to the effective dielectric constant $\epsilon$. The optimized ground-state energy is plotted as a function of the bare Coulomb scale $U_0=\epsilon U$ for $\epsilon=8$--$12$, with $t_2/t_1=0$, $t_3/t_1=0.20$, and $J/t_1=0.25$ at half filling. The dashed horizontal line denotes the energy of the Mott solution with $g\rightarrow0$. Changing $\epsilon$ shifts the bare interaction scale at which the SC--Mott crossing occurs but preserves the intermediate gossamer-SC regime.
    }
    \label{fig:epsilon_robustness}
\end{figure}

\section{E. Estimation of $T_c$ and Coherence Length}\label{app:Tc_xi}

The RMFT provides a variational estimate of the ground-state energy but does not directly determine the true superconducting transition temperature $T_c$ or the coherence length $\xi$, which are governed by phase fluctuations beyond the mean-field level. Nevertheless, one may still obtain order-of-magnitude estimates by combining the RMFT-optimized parameters with the standard weak-coupling BCS theory and the Ginzburg--Landau formalism.

The Gutzwiller projection suppresses double occupancies and renormalizes the pairing amplitude. The physical pairing gap is related to the mean-field gap by~\cite{Zhang2003}
\begin{equation}
    \Delta_{\rm phys} \approx g^{2}\,\overline{\Delta}_{\rm MF, FS},
    \label{eq:delta_phys}
\end{equation}
where $g$ is the optimized Gutzwiller factor and $\overline{\Delta}_{\rm MF, FS}$ is the mean-field $d+id$ pairing amplitude averaged over the Fermi surface.

While estimation of transition temperature within any mean field theory, including the renormalized mean field theory would need special caution. We could very roughly also estimate $T_{c}$ by using the renormalized SC gap. The mean-field $T_c$ is estimated from the weak-coupling BCS relation,
\begin{equation}
    k_{B}T_{c}^{\rm RMFT} \approx \frac{\Delta_{\rm phys}}{2.14}.
    \label{eq:Tc_bcs}
\end{equation}
For the superconducting regime $U/t_{1}=8$, $t_{3}/t_{1}=0.2$, and $J/t_{1}=0.25$, we find $g\approx0.57$, $\overline{\Delta}_{\rm MF, FS}\approx0.084\,t_{1}$. With $t_{1}\approx2.4$\,meV from Wannierization, $\Delta_{\rm phys}\approx0.07$\,meV and $T_{c}^{\rm RMFT}\approx0.4$\,K. This is an upper bound, as BKT phase fluctuations in two dimensions will further reduce the observed $T_c$. The estimate is consistent in order of magnitude with the experimental $T_c\sim0.2$\,K~\cite{Xia2026}.

We further estimate the the coherence length $\xi$ from the Ginzburg--Landau theory, which is related to the Fermi velocity and the physical pairing gap by
\begin{equation}
    \xi \sim \frac{v_{F,{\rm phys}}}{\pi|\Delta_{\rm phys}|}
    = \frac{v_{F,{\rm MF}}}{\pi|\overline{\Delta}_{\rm MF, FS}|},
    \label{eq:xi}
\end{equation}
where $v_{F,{\rm MF}}$ is the Fermi-surface-averaged Fermi velocity of the optimized mean-field state. The second equality follows because both the kinetic energy and the pairing amplitude are renormalized by the same Gutzwiller factor $g^{2}$ to leading order [see Eq.~(\ref{eq:delta_phys}) and the discussion of the hopping renormalization in Sec.~B], so that $g^{2}$ cancels in the ratio $v_{F,{\rm phys}}/|\Delta_{\rm phys}|$. Consequently, at this level of approximation, $\xi$ is not directly renormalized by the suppression of charge fluctuations. For the superconducting regime $U/t_{1}=8$, $t_{3}/t_{1}=0.2$, and $J/t_{1}=0.25$, we obtain $\xi\approx7\,a_{M}$, which is comparable in order of magnitude to the experimentally reported value $\xi\approx8\,a_{M}$~\cite{Xia2026}. Qualitatively speaking, SC coherence length is short in gossamer superconductor due to electron strong correlation, similar to the doped Mott insulator such as in high $T_c$ cuprates. The experimental value of 8 times of moire lattice appears to be consistent with this picture. 

\begin{figure}[!t]
    \centering
    \includegraphics[width=0.55\linewidth]{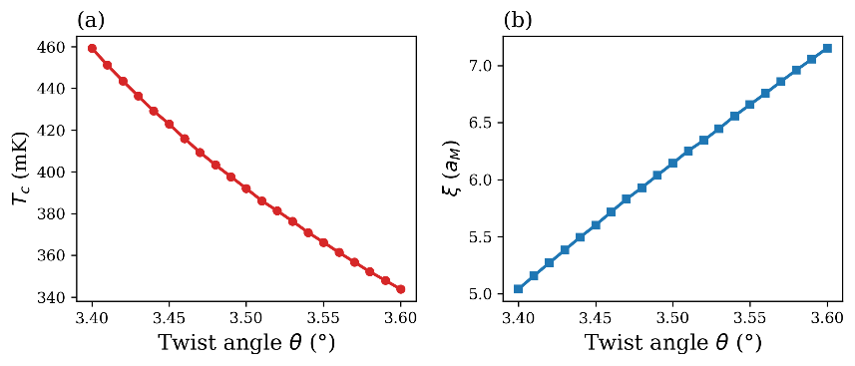}
    \caption{
    Estimated $T_{c}$ and coherence length $\xi$ along the gossamer SC phase boundary at half-filling.
    (a) RMFT $T_{c}$ from $k_{B}T_{c}\approx0.47\,g^{2}\overline{\Delta}_{\rm MF, FS}$.
    (b) Coherence length $\xi=v_{F,{\rm MF}}/(\pi\overline{\Delta}_{\rm MF, FS})$.
    The effective $U/t_{1}$ increases from $8.2$ at $\theta=3.60^{\circ}$ to $9.5$ at $\theta=3.40^{\circ}$.
    }
    \label{fig:tc_xi_vs_theta}
\end{figure}

Figure~\ref{fig:tc_xi_vs_theta} shows the RMFT $T_{c}$ and $\xi$ as a function of twist angle in the range $\theta=3.40^{\circ}$--$3.60^{\circ}$, where the variation is captured by the change in $U/t_{1}$ (from $9.5$ to $8.2$). Other parameters are fixed as $t_{2}/t_{1}=0$, $J/t_{1}=0.25$, $t_{3}/t_{1}\approx0.19$. We find that $T_{c}$ decreases and $\xi$  increases as the twist angle increases. Reducing $U/t_{1}$ (increasing $\theta$) weakens the effect of $J$, which suppresses the pairing gap and raises the superfluid stiffness, thereby lowering $T_{c}$. The estimated $T_{c}$ values are consistent in order of magnitude with experiment~\cite{Xia2026}, with the discrepancy attributable to BKT fluctuations not captured in the mean-field treatment.
\end{widetext}

\end{document}